\pdfoutput=1
\documentclass[twocolumn]{aastex7}
\usepackage{amsmath}
\usepackage{verbatim}
\usepackage{rotating}
\usepackage{makecell}
\usepackage{tabularx}
\usepackage{array}
\usepackage{subfig}
\usepackage{graphicx}
\usepackage{booktabs}
\usepackage{multirow}
\usepackage{parcolumns}
\usepackage{afterpage}
\usepackage{placeins}

\begin{document}

\title{A Large Diversity in the Quiescent Fractions of Massive Galaxies in Proto-clusters at $z\sim4-6$ in the FRESCO GOODS fields}

\author[0000-0002-3101-8348]{Jamie Lin}
\affiliation{Department of Physics and Astronomy, Tufts University, 574 Boston Avenue, Suite 304, Medford, MA 02155, USA}
\email{Jamie.Lin@tufts.edu}

\author[0000-0001-9002-3502]{Danilo Marchesini}
\affiliation{Department of Physics and Astronomy, Tufts University, 574 Boston Avenue, Suite 304, Medford, MA 02155, USA}
\email{Danilo.Marchesini@tufts.edu}

\author[0000-0001-5851-6649]{Pascal A. Oesch}
\affiliation{Department of Astronomy, University of Geneva, Chemin Pegasi 51, 1290 Versoix, Switzerland}
\affiliation{Cosmic Dawn Center (DAWN), Denmark}
\affiliation{Niels Bohr Institute, University of Copenhagen, Jagtvej 128, 2200 Copenhagen, Denmark}
\email{pascal.oesch@unige.ch}

\author[0000-0003-3997-5705]{Rohan P. Naidu}
\affiliation{MIT Kavli Institute for Astrophysics and Space Research, 77 Massachusetts
Avenue, Cambridge, MA 02139, USA}
\email{rnaidu@mit.edu}

\author[0000-0002-9672-3005]{Alba Covelo-Paz}
\affiliation{Department of Astronomy, University of Geneva, Chemin Pegasi 51, 1290 Versoix, Switzerland}
\email{Alba.Covelopaz@unige.ch}

\author[0000-0003-2680-005X]{Gabriel Brammer}
\affiliation{Cosmic Dawn Center (DAWN), Denmark}
\affiliation{Niels Bohr Institute, University of Copenhagen, Jagtvej 128, 2200 Copenhagen, Denmark}
\email{gabriel.brammer@nbi.ku.dk}

\author[0000-0003-2164-7949]{Thomas Herard-Demanche}
\affiliation{Leiden Observatory, Leiden University, P.O. Box 9513, 2300 RA Leiden, The Netherlands}
\email{herard@strw.leidenuniv.nl}

\author[0000-0002-8896-6496]{Christian Kragh Jespersen}
\affiliation{Department of Astrophysical Sciences, Princeton University, Princeton, NJ 08544, USA}
\email{cj1223@princeton.edu}

\author[0000-0001-8928-4465]{Andrea Weibel}
\affiliation{Department of Astronomy, University of Geneva, Chemin Pegasi 51, 1290 Versoix, Switzerland}
\email{Andrea.Weibel@unige.ch}

\author[0000-0001-7160-3632]{Kate Whitaker}
\affiliation{Department of Astronomy, University of Massachusetts, Amherst, 710
N Pleasant Street, Amherst, MA, 01003, USA.}
\affiliation{Cosmic Dawn Center (DAWN), Denmark}
\email{kwhitaker@astro.umass.edu}

\author[0000-0003-1207-5344]{Mengyuan Xiao}
\affiliation{Department of Astronomy, University of Geneva, Chemin Pegasi 51, 1290 Versoix, Switzerland}
\email{Mengyuan.Xiao@unige.ch}

\begin{abstract}
To constrain the emergence of the first quiescent galaxies (QGs) and their connection to large-scale structure, we search for rare, high-redshift QGs in proto-clusters using spectral energy distribution modeling, multiple rest-frame color diagnostics, and visual inspection. Our sample includes 12 QG candidates with stellar masses $\log(M_*/M_{\odot}) \sim 9-11$ in four proto-clusters at $z\sim4-6$ within GOODS-North (GOODS-N) and GOODS-South (GOODS-S). The selection uses 11 NIRCam filters in GOODS-N and 17 NIRCam plus 2 MIRI filters in GOODS-S, combined with 9 ancillary HST filters, covering $0.4$--$7.7\,\mu$m. We measure an increasing quiescent fraction with stellar mass, with averages of $1.0\%^{+0.7}_{-0.4}$ (5/502), $3.5\%^{+2.8}_{-1.7}$ (4/115), and $33.3\%^{+22.2}_{-17.5}$ (3/9) for $\log$(M$_{*}$/M$_{\odot}$) = 9 -- 9.75, 9.75 -- 10.5, 10.5 -- 11.25, respectively, across four proto-clusters. Notably, the two GOODS-N proto-clusters at $z=4.406$ and $z=5.194$ have a quiescent fraction of $100\%$ (2/2 at $z=4.406$; 1/1 at $z=5.194$) in their highest mass bin ($\log$(M$_{*}$/M$_{\odot}$) $> 10.5$), whereas the proto-clusters in GOODS-S at similar redshifts have zero massive quiescent galaxies. Contrary to the expectation that massive galaxies occupy cluster centers at $z\sim0$, our QGs are found primarily on the outskirts of overdensity peaks. We also find fewer robust QGs at $z>5$ (1/8) and none in the $z=5.384$ proto-cluster, consistent with simulations suggesting that efficient quenching mechanisms had not yet matured. Our results highlight the pronounced variations in quiescent fractions across proto-clusters at $z>4$. The emerging diversity of proto-cluster QG populations calls for a stronger understanding of the role of environment in early galaxy evolution.
\end{abstract}


\section{Introduction}
\label{sec:intro}

How and when the first massive galaxies form and subsequently shut down their star formation remains one of the major puzzles of galaxy evolution. Recent James Webb Space Telescope (JWST) observations reveal massive quiescent galaxies at $z>4$ and ancient quiescent galaxies at $z\sim3$ that may have formed most of their stars by $z\sim11$ \citep{Carnall2023a, Carnall2023b, Glazebrook2024, deGraaff2024, Weibel2025}. Their high-redshift nature implies rapid mass assembly and quenching within the first two billion years after the Big Bang. Moreover, many of these massive quiescent galaxies have been found to reside in proto-clusters \citep{deGraaff2024, Forrest2024a, McConachie2022}. This raises questions about the abundance of massive quiescent galaxies in high-redshift proto-clusters, 
and the role of environment in shaping galaxy evolution during the early universe.


In the local universe, many massive quiescent galaxies (e.g., ellipticals or S0s) reside in dense cluster environments and occupy the peaks of overdensities \citep{Thomas2011}. These galaxies experienced enhanced star-formation activity early in their histories compared to field galaxies, allowing them to grow early and evolve rapidly. At higher redshifts, several studies have also reported elevated quiescent fractions ($f_{QG}$) in proto-clusters at $z>2-4$, ranging from $\sim20\%$ at $z = 2 - 2.5$, $\sim30\%$ at $z = 3.3$, $\sim70\%$ at $z = 3.37$, and $\sim 40\%$ at $z = 4$ \citep{Edward2024, Forrest2024a, McConachie2022, Tanaka2024}. However, the measured quiescent fractions at $z\sim2-4$ still vary considerably. With the ability to detect both high redshift proto-clusters and quiescent galaxies with JWST, we can now begin to constrain the fraction of quiescent galaxies residing in proto-clusters at $z>4$. 

There are several established approaches for identifying high-redshift proto-clusters. These include detecting Ly$\alpha$ emitters associated with intense star formation in the early universe, searching for radio galaxies and quasars that are often located near the centers of massive halos \citep{Postman2000}, or using narrow- and medium-band filters with central wavelengths specifically matched to prominent emission lines from proto-cluster members \citep{Cooke2014, Muldrew2015}. In this work, we leverage a method that has proven particularly effective with JWST wide-field slit-less spectroscopy: identifying spatial overdensities of emission-line galaxies within a tightly constrained redshift interval.

JWST wide-field slit-less spectroscopy enables the detection and spectroscopic redshift confirmation of line-emitting galaxies over a field of view of 2.2' $\times$ 2.2' per pointing. The F356W and F444W  grism observations cover wavelength ranges of 3.14-3.98 $\mu$m and 3.88-4.99 $\mu$m, respectively. By combining these two bands, a census of H$\alpha$ and [OIII] emitters at $z\sim4-6.6$ and $z\sim6.7-9$ can be performed. 

Numerous proto-clusters have been identified using H$\alpha$ emitters in the JWST FRESCO program in the GOODS-N and GOODS-S fields \citep{Herard-Demanche2023, Helton2024} and in the JWST CONGRESS program in GOODS-N \citep{Alba2014} at $z\sim4-5$. In this study, we search for quiescent galaxies within these proto-clusters in the GOODS fields at $z\sim4-6$. Our goals are to assess whether quiescent galaxies are present in proto-clusters at these redshifts and to measure the fraction of quiescent galaxies in these early extreme environments. 

The structure of this paper is as follows. In Section \ref{sec:data}, we introduce the JWST FRESCO survey and ancillary datasets used in this work. In Section \ref{sec:method}, we discuss our spectral energy distribution (SED) fitting methodology and quiescent sample selection. In Section \ref{sec:results},  we present our results, including overdensity maps and quiescent fractions at $z\sim4-6$. We discuss our findings in Section \ref{sec:discussion}. 
For cosmological calculations, we adopt a $\Lambda$CDM cosmology with $\Omega_{M}$= 0.3, $\Omega_{\Lambda}$= 0.7, H$_0$=70 km s$^{-1}$ Mpc$^{-1}$, and we assume a \cite{Kroupa2001} initial mass function. All magnitudes are quoted in the AB magnitude system \citep{Oke1983}.

\section{Data} \label{sec:data}
For our analysis, we primarily use data from the First Reionization Epoch Spectroscopically Complete Observations (FRESCO; \citealt{Oesch2023}) JWST Cycle 1 medium program (GO-1895; 53.8 hr). The FRESCO survey covers the two well-studied GOODS/CANDELS fields, GOODS-N and GOODS-S, for a total area of 124 arcmin$^2$. In the program, we obtained $\sim 2$ hr deep NIRCam/grism observations with the F444W filter and a mosaic of F182M, F210M, and F444W imaging in the same fields to a 5$\sigma$ depth of $\sim$ 28.2 mag. In this work, we focus on sources with reliable JWST/NIRCam detections (i.e., \verb|use_phot=1|\footnote{Flag from \citet{Weibel2024} to remove bright stars and their spikes and to mask out noisy regions. See \citet{Weibel2024} for details on how \texttt{use\_phot=1} is defined.}  and S/N $>$ 3 in F444W). For more information on image data reduction and the photometric catalogs, refer to \citet{Weibel2024}. 

We combine ancillary HST/ACS and WFC3/IR data with NIRCam and MIRI data from other JWST programs, namely JADES \citep{Eisenstein2023} and JEMS \citep{Williams2023},  to model the spectral energy distribution (SED) of galaxies using the software BAGPIPES \citep{Carnall2018}. The SEDs are sampled by 11 NIRCam filters in GOODS-N, 17 NIRCam and 2 MIRI filters in GOODS-S, combined with 9 HST filters, together covering observed wavelengths of $\sim$ 0.4 - 4.4 $\mu m$ in GOODS-N and $\sim$ 0.4 - 7.7 $\mu m$ in GOODS-S. We adopt the 16th and 84th percentiles of photometric redshift p(z) from EAZY \citep{Brammer2008} in the photometric catalogs as the prior for our SED models \citep{Weibel2024}. For more details about assumptions and models used for SED modeling with BAGPIPES, refer to Section \ref{sec:sed}.

Lastly, we use JWST/NIRCam slit-less spectroscopy in filter F444W from the FRESCO survey, covering $\lambda = 3.9 - 5.0 \mu$m, and F356W from the Complete NIRCam Grism Redshift Survey (CONGRESS; Cycle 2, \citealt{congress2023}), covering $\lambda = 3.1 - 4.0 \mu$m, to check for a lack of rest-frame optical emission lines (e.g., H$\alpha$+[NII], [OIII]+H$\beta$) confirming low star-formation activity for our quiescent sample and to identify proto-clusters using H$\alpha$ emitting galaxies. 

All the JWST data used in this paper can be found in MAST: \dataset[https://doi.org/10.17909/axfr-9n93]{https://doi.org/10.17909/axfr-9n93}.

\section{Method} \label{sec:method}

\subsection{SED Models} \label{sec:sed}
To derive physical properties of galaxies from observed photometric data, we employ the spectral energy distribution modeling code BAGPIPES \citep{Carnall2018}. We incorporate the simple stellar population (SSP) synthesis models from \citet{Bruzual2003}, the \citet{Kroupa2001} initial mass function (IMF), and the \citet{Calzetti2000} dust attenuation curve in BAGPIPES. We also include a nebular component, which accounts for emission lines and nebular continuum emission, modeled with CLOUDY \citep{ferland2013}.

We assume a double-power-law star-formation history (SFH) model and apply the 16th to 84th percentile range of EAZY's redshift probability distribution function $p(z)$ as a prior for redshift. EAZY's redshift is  used only as a prior for our SED fitting. All of our analyses that involve photometric redshift selection utilize the best-fit redshift from BAGPIPES unless otherwise specified. For galaxies with spectroscopic redshifts, SED fitting is performed at fixed spectroscopic redshift. 

We use the best-fit parameters (i.e., redshift $z$, stellar mass M$_{*}$, star-formation rate (SFR), specific star-formation rate (sSFR), and mass-weighted age (Age$_{mw}$)) from BAGPIPES to select our quiescent sample. We present the priors adopted for the model parameters in Table \ref{tab:sed}.

\begin{deluxetable}{lccc}[!ht]
\tabletypesize{\scriptsize} 
\tablewidth{0pt} 
\tablecaption{Parameter settings for \texttt{BAGPIPES}\label{tab:sed}}
\tablehead{
\colhead{Parameter} &
\colhead{Unit} &
\colhead{Range} &
\colhead{Prior}
}
\startdata
\multicolumn{4}{c}{\textbf{General}$^{a}$} \\[2pt]
Stellar Mass Formed & $\log(M_\star/M_\odot)$ & (5, 15) & Logarithmic \\
Peak Star Formation & $\tau$ (Gyr) & (0.03, 10) & Uniform \\
Age & $t_{\rm age}$ (Gyr) & (0.001, $t_{\rm obs}$) & Logarithmic \\
Metallicity & $Z/Z_\odot$ & (0.04, 2.5) & Uniform \\
V-band attenuation & $A_V$ & (0, 6) & Uniform \\
Birth-cloud factor & $\eta$ & (0, 3) & Uniform \\
Ionization & $U$ & ($-4$, $-2$) & Logarithmic \\
\hline
\multicolumn{4}{c}{\textbf{Double-power-law}$^{b}$} \\[2pt]
Falling slope & $\alpha$ & (0.01, 1000) & Logarithmic \\
Rising slope & $\beta$ & (0.01, 1000) & Logarithmic \\
\enddata
\tablecomments{
$^{a}$ For galaxies with photometric redshifts, the redshift prior is defined by the 16th and 84th percentiles of the EAZY $p(z)$. For galaxies with spectroscopic redshifts, the redshift is fixed. \\
$^{b}$ Parameters for the double-power-law star-formation history model.
}
\end{deluxetable}

\subsection{Rest-frame colors} \label{sec:colors}
After performing SED fitting, we derive the rest-frame colors, $NUV-U$, $U-V$, $V-J$\footnote{The derivation of rest-frame $U-V$ and $V-J$ is described in Section \ref{sec:colors}. The rest-frame J is not directly constrained by the observed bandpass at these redshifts. As a result, colors involving J (e.g., V - J) are effectively SED-model outputs rather than purely data-driven measurements.}, and the synthetic colors $u_s$ - $g_s$ and $g_s$ - $i_s$ from \citet{Antwi-Danso2023}, from the posterior distributions of the model SEDs to select quiescent galaxies. To derive the best-fit \textit{UVJ} colors for each galaxy, we compute U-V and V-J, using the Maíz Apellániz U and V filters and the 2MASS J filters, for each sample in the posterior, and take the median of these color distributions as the best-fit values. The uncertainties are estimated from the 16th and 84th percentiles of the U-V and V-J distributions, representing the 1$\sigma$ lower and upper errors, respectively. 

To derive the $(ugi)_{s}$ and NUVU-VJ colors, we use the posterior distributions of the model SEDs to generate those for the rest-frame $u_s$, $g_s$, $i_s$, U, V, J, and NUV magnitudes using sed.py \citep{sedpy}. We first shift the SED model posterior distribution of each galaxy into the rest frame using the best-fit redshift from BAGPIPES. Then, we calculate AB magnitudes for a given SED and photometric filter curve. Finally, we compute the colors $u_s$-$g_s$, $g_s$-$i_s$, NUV-U, and V-J from the corresponding magnitude values in each sample within the posterior. The median of the resulting color distribution is adopted as our best-fit color, and the 1$\sigma$ uncertainties are estimated using the 16th and 84th percentiles of the color distribution.

\subsection{Proto-clusters at $z\sim4.25-4.40$ and $z\sim5.20-5.40$ in GOODS-N and GOODS-S} \label{sec:overdensity}

Significant high-redshift galaxy overdensities at $z\sim4-6$ have been reported in recent H$\alpha$ emission line studies in GOODS-N and GOODS-S fields, using JWST/NIRCam imaging and wide-field slit-less spectroscopy from the FRESCO and CONGRESS surveys \citep{Herard-Demanche2023,Helton2024, Alba2014}. Specifically, \citet{Alba2014} identified a massive structure in GOODS-N at $z=4.43\pm0.03$ encompassing 97 H$\alpha$ emitters. At higher redshifts, \citet{Herard-Demanche2023} revealed 100 H$\alpha$ emitters across 10 sub-structures at $z=5.17–5.20$    within a volume of approximately $\sim15(cMpc)^3$. In the GOODS-S field, \citet{Helton2024} mapped an overdensity at $z\sim5.4$ consisting of 96 H$\alpha$ emitters, 42 of which are spectroscopically confirmed members. 

\begin{figure}
    \includegraphics[width=.45\textwidth]
    {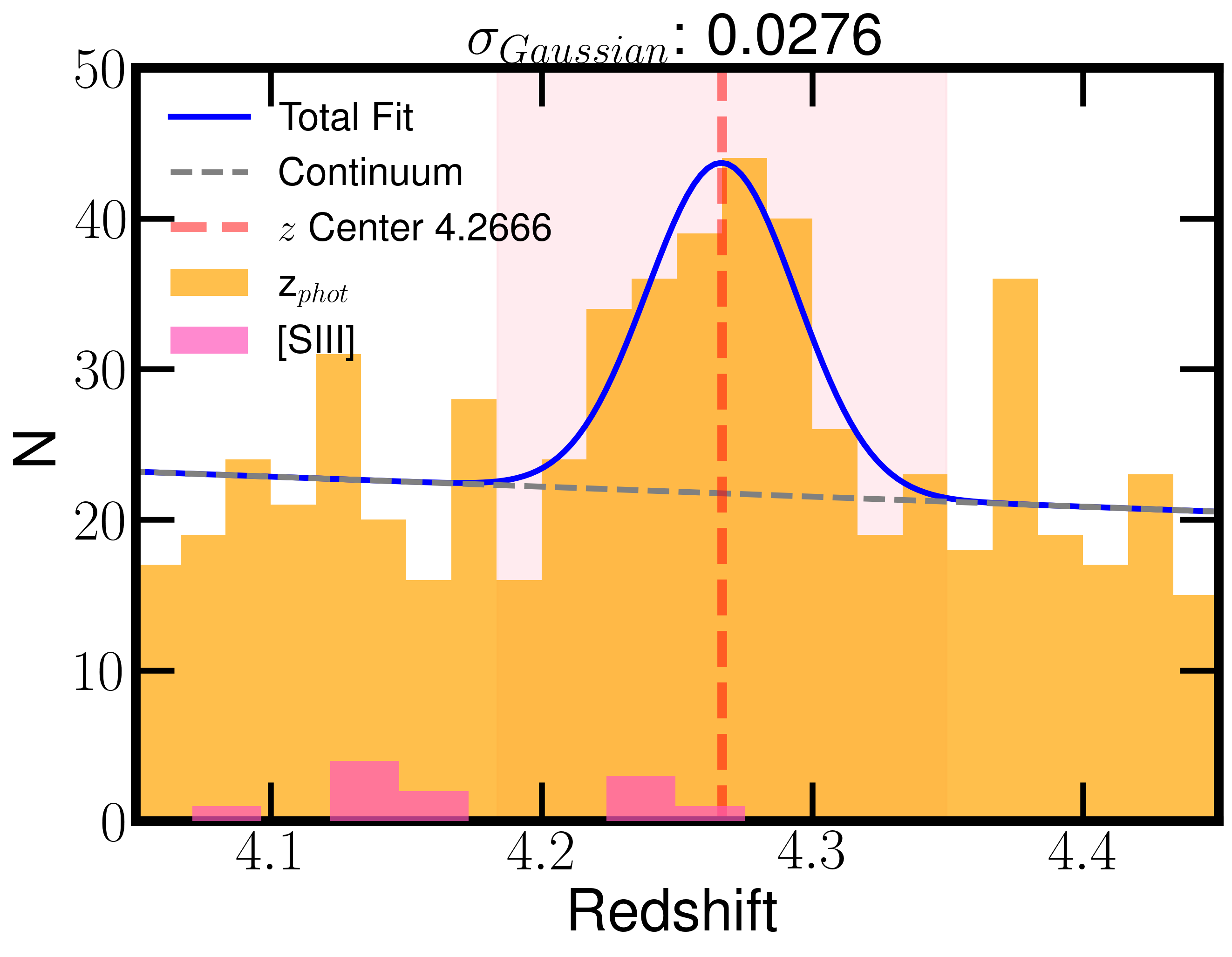}
    \caption{Photometric redshift distribution of sources around the proto-cluster at $z_{od} = 4.267$ in GOODS-S. A prominent peak is visible and fitted with a Gaussian plus continuum model. The total fit is shown as a blue curve, whereas the gray line shows the continuum. The center of the Gaussian fit is defined as the redshift of the proto-cluster $z_{od} = 4.267$ (red dashed line). The width of $\pm 3\times\sigma_{Gaussian}$ is shown as the pink shaded region centered at the $z_{od} = 4.267$. We also show the spectroscopic redshift distribution from [SIII] lines (pink) in FRESCO. However, the handful of robust [SIII] detections is insufficient to confirm the overdensity.}
    \label{fig:gds-z4.3-zdistribution}
\end{figure}

In this work, we complement these three spectroscopic structures by identifying a new candidate overdensity at $z\sim4.3$ in GOODS-S using photometric redshifts derived from our SED modeling of the FRESCO catalogs. As shown in Figure \ref{fig:gds-z4.3-zdistribution}, we find a prominent overdensity of 217 galaxies concentrated within $z_{phot} \sim 4.2-4.3$. Because the GOODS-S footprint lacks matching F356W grism spectroscopy, we cannot confirm this structure via typical strong rest-frame optical lines like H$\alpha$ or [OIII]. An examination of the weaker [SIII]$\lambda9532$ line within the F444W grism data yields only a handful of robust line detections, which are insufficient for full spectroscopic validation. Consequently, we rely primarily on photometric tracking for this single structure. In total, our final analysis tracks four independent overdensities, summarized spatially in Figures \ref{fig:gdn-overdensity-all} and \ref{fig:gds-overdensity-all} in Section \ref{sec:results}.

\begin{deluxetable}{lccc}
    \tabletypesize{\normalsize}
    \setlength{\tabcolsep}{10pt}
    \tablecaption{Photometric Redshift Selection Windows for Target Overdensities \label{table:overdensity-zphot}}
    \tablehead{
    \colhead{Field} &
    \colhead{$z_{\rm od, spec}$} &
    \colhead{$z_{\rm od, phot}$} &
    \colhead{$z_{\rm phot}$}
    }
    \startdata
    GOODS--S & N/A & $4.267$ & 4.10 -- 4.40 \\ 
    GOODS--N & $4.406$ & $4.456$ & 4.20 -- 4.70 \\ 
    GOODS--N & $5.194$ & $5.197$ & 5.00 -- 5.40 \\ 
    GOODS--S & $5.384$ & $5.452$ & 5.30 -- 5.60 \\ 
    \enddata
    \tablecomments{Spectroscopic overdensity centers ($z_{\rm od}$), offset-corrected photometric overdensity centers ($z_{\rm od, phot}$) and corresponding selection slices ($z_{\rm phot}$) used to select cluster members. For the three spectroscopically confirmed overdensities, boundaries correspond to $z_{\rm od, phot} \pm 2\sigma(\Delta z)$ derived from local $\sigma_{\rm NMAD}$ scatter. For the purely photometric overdensity at $z=4.267$, the window matches a $\pm 3\sigma_{\rm Gaussian}$ width that is roughly equal to $\sim 2\sigma(\Delta z)$ in the three spectroscopically confirmed overdensities. The upper and lower bounds are rounded outward to the nearest 0.1, maximizing completeness of the proto-clusters' members. }
\end{deluxetable}

To reliably search for quiescent galaxies embedded within these four environments, we must establish rigorous photometric redshift windows ($z_{phot}$ ranges) that maximize completeness while minimizing field contamination. For the three overdensities with spectroscopic confirmation, we first calculate the typical scatter between photometric and spectroscopic redshifts using all available high-quality (redshift quality flag q$>$1.75) spectroscopic catalog sources. We calculate the normalized median absolute deviation ($\sigma_{\mathrm{NMAD}}$
 ) via:

\begin{equation}
\sigma_{\rm NMAD} = 1.48 \times {\rm median}\left(\frac{|\Delta z - {\rm median}(\Delta z)|}{1 + z_{\rm spec}}\right)
\label{eq:nmad}
\end{equation}

where $\Delta z = z_{phot} - z_{spec}$. We multiply $\sigma_{\mathrm{NMAD}}$ by (1 + $z_{od}$) to get $\sigma(\Delta z)$. Using the robust emission-line samples, we find $\sigma(\Delta z) \simeq 0.1194$ centered at the $z_{od} = 4.406$ peak (based on 534 emitters), $\sigma(\Delta z) \simeq 0.0780$ at the $z_{od} = 5.194$ (271 emitters), and $\sigma(\Delta z) \simeq 0.0713$ at the $z_{od} = 5.384$ (67 emitters).

We define the redshift of the proto-clusters as $z_{od} =$ 4.406, 5.194, and 5.384 from the median of $z_{spec}$ around the peak (i.e., $4.4\pm 0.1$, $5.2\pm0.1$, and $5.4\pm0.1$). To define the photometric selection windows, we first measure the median systematic offset ($\Delta z_{\mathrm{offset}}$) between the $z_{spec} \pm 0.1$ peak and the corresponding $z_{phot}$ peak location for each structure. The expected photometric center of the overdensity is then defined as  $z_{od,phot} = z_{od} +$ $\Delta z_{\mathrm{offset}}$.  We then establish a conservative search window spanning $z_{od,phot} \pm 2\sigma(\Delta z)$.

For our newly identified photometric overdensity z$\sim$4.3, an alternative statistical approach is required due to the absence of a large spectroscopic sample. We fit a composite Gaussian plus continuum model directly to the photometric redshift distribution shown in Figure \ref{fig:gds-z4.3-zdistribution}. The best-fit Gaussian centroid defines the photometric center at $z_{od,phot} = 4.267$, with an intrinsic width of $\sigma_{Gaussian} = 0.0276$. For this structure, we define the member selection window using a $\pm3 \sigma_{Gaussian}$ width. This boundary definition is chosen deliberately because we compared the $\sigma_{Gaussian}$ width with $2\sigma(\Delta z)$ of the three spectroscopically identified proto-clusters and found that $3\sigma_{Gaussian}$ roughly corresponds to $2\sigma(\Delta z)$.

We select quiescent candidates whose BAGPIPES-derived redshift distributions physically intersect with the target windows. Specifically, a galaxy is included as a member if its median photometric redshift, or its associated 1$\sigma$  interval (bounded by the 16th and 84th percentiles of the posterior probability distribution), falls within the defined range for that overdensity. Finally, to maximize sample completeness within each overdensity, the upper and lower bounds of the z phot windows based on $2\sigma(\Delta z)$ or $3\sigma_{Gaussian}$ are rounded outward to the nearest 0.1. The final optimized photometric redshift selection ranges used for the four proto-clusters are summarized in Table \ref{table:overdensity-zphot}.

Using the stellar-mass-halo mass relation from \citet{Behroozi2019}, we first estimate the dark matter halo masses of the $z=4.267$ (GS), $z=4.406$ (GN), $z=5.194$ (GN), $z=5.384$ (GS) proto-clusters using the observed stellar masses of both spectroscopic and photometric members. This yields $\log$($M_{h}/M_{\odot}$) of $13.87$, $14.09$, $13.96$, $13.46$, respectively. These initial estimates are higher than those of other spectroscopically confirmed proto-clusters such as SPT 2349-56 at $z\sim$ 4.3 (10$^{13.48}$); \citet{Chapman2024}, and JADES-ID1 at ~$z \sim 5.7$ (10$^{13.28}$; \citet{Bogdan2026}). This is likely driven by our photometric selection, which was optimized to maximize the completeness of quiescent candidates while introducing field contaminants into the proto-cluster members. To mitigate this effect, we recalculate the halo masses using only spectroscopic members of the proto-clusters. Because the $z=4.267$ (GS) overdensity lacks a spectroscopic sample (only one candidate member, GS-21434, has a spectroscopic redshift; Section \ref{sec:dja}), it is excluded from this subset. For the remaining structures, we estimate $\log$($M_{h}/M_{\odot}$) of $13.63$ ($z=4.406$, GN), $13.65$ ($z=5.194$, GN), and $12.98$ ($z=5.384$, GS). These values are much more comparable to literature values. Consequently, the true dark matter halo masses of these proto-clusters are likely bracketed by these two boundaries, with the spectroscopic-only and combined totals serving as lower and upper limits, respectively.

\subsection{Quiescent Galaxy Selection} \label{sec:quiescent-selection}
We make use of multiple methods to select quiescent candidates in the four proto-clusters in the FRESCO GOODS-N and GOODS-S fields at $z\sim4-6$ to mitigate the biases of each selection method:
\begin{itemize}
    \item UVJ \citep{Schreiber2015, Belli2019}
    \item (ugi)$_s$ \citep{Antwi-Danso2023}
    \item GMM/NUVU$-$VJ \citep{Gould2023}
    \item sSFR/yr $\leq$ $0.2/t_{obs}$ 
\end{itemize}

We will discuss each method in detail in the following subsections (Section \ref{sec:uvj} -- \ref{sec:low-ssfr}). For the GMM/NUVU-VJ method, we adopt two probability thresholds, p(q) $>$ 0.1 and p(q) $>$ 0.01 (Section 3.4.3), giving five selection criteria in total (M1–M5 in Table \ref{table:quiescent-candidates}). All galaxies in our final sample satisfy the UVJ and sSFR criteria and pass visual inspection. If a source passes all five criteria, we consider that galaxy a "robust" quiescent candidate, and the remaining candidates are considered less robust.

\begin{figure*}
    \centering
    
    \vspace{-2.0em}
    \includegraphics[width=0.4\textwidth]{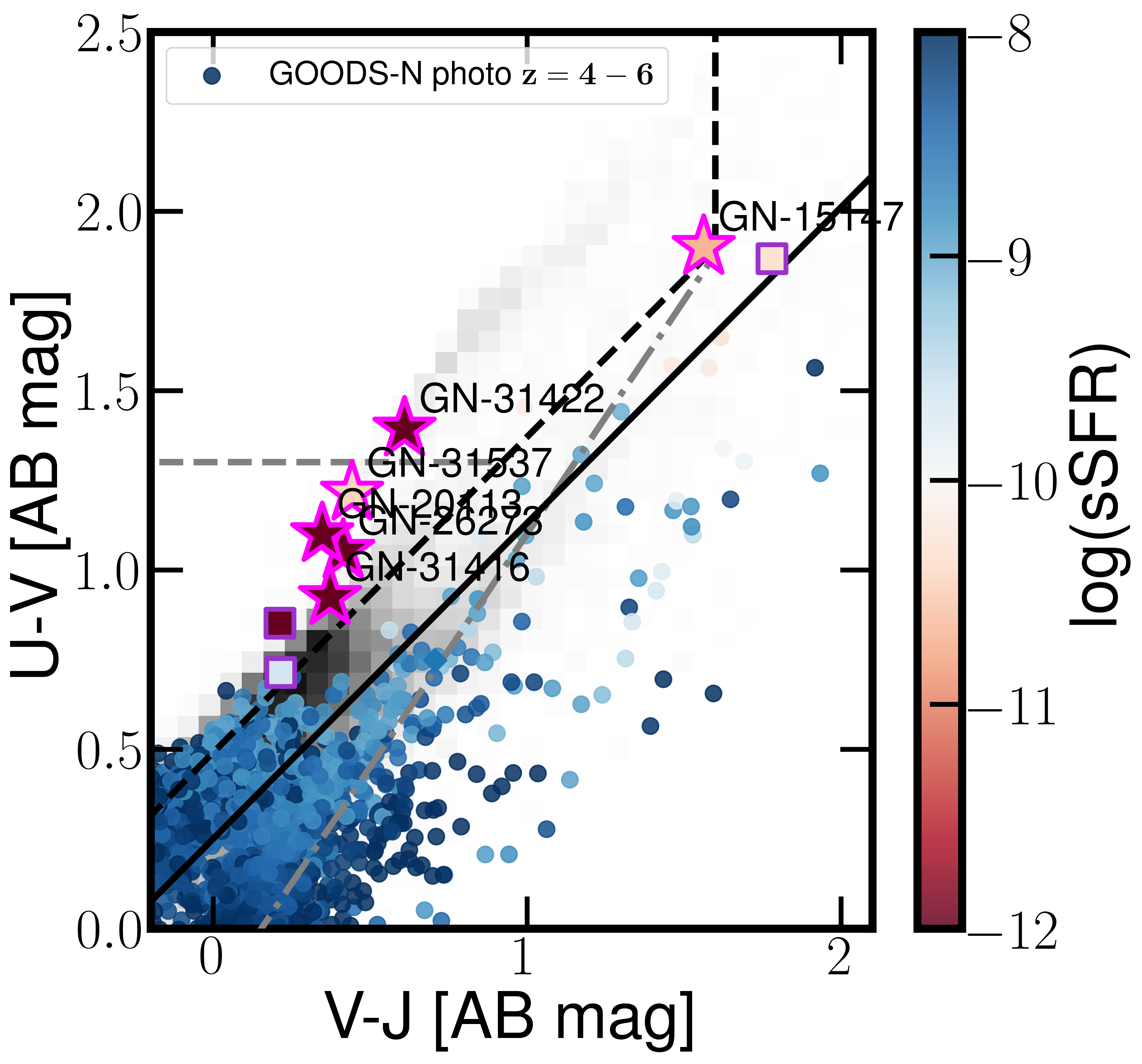}
    \hspace{0.01\textwidth}
    \includegraphics[width=0.4\textwidth]{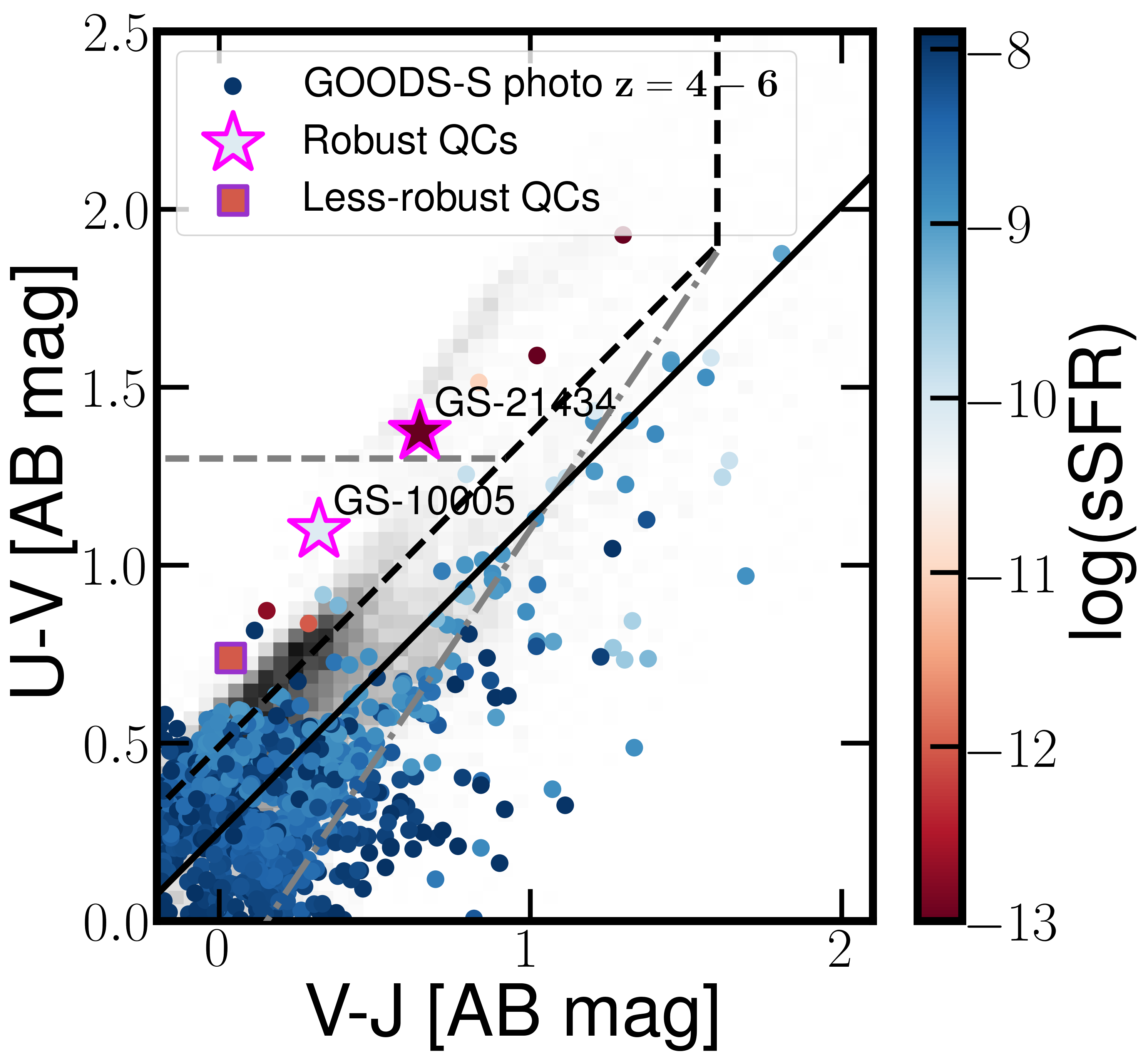}\\

    \includegraphics[width=0.4\textwidth]{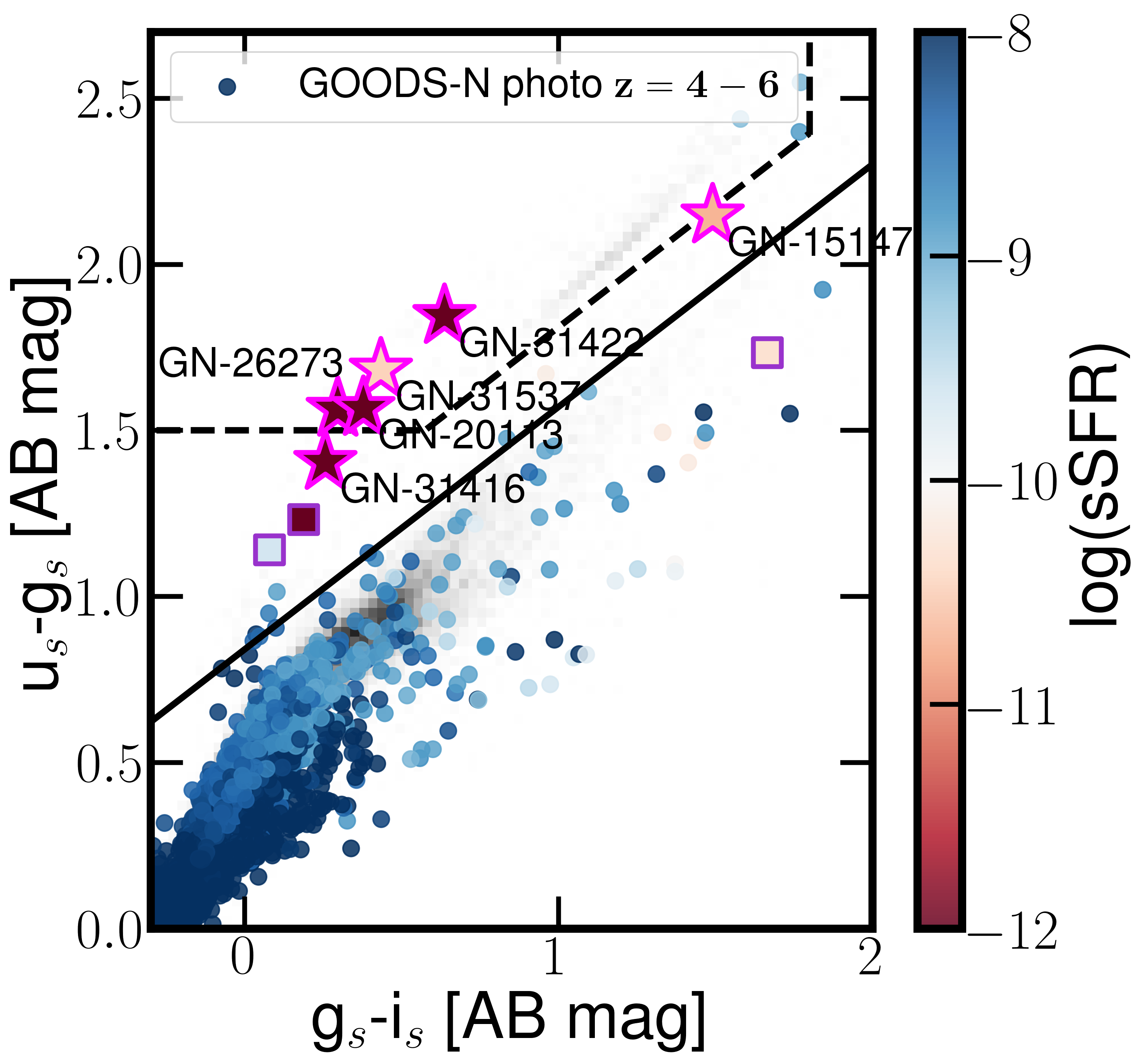}
    \hspace{0.01\textwidth}
    \includegraphics[width=0.4\textwidth]{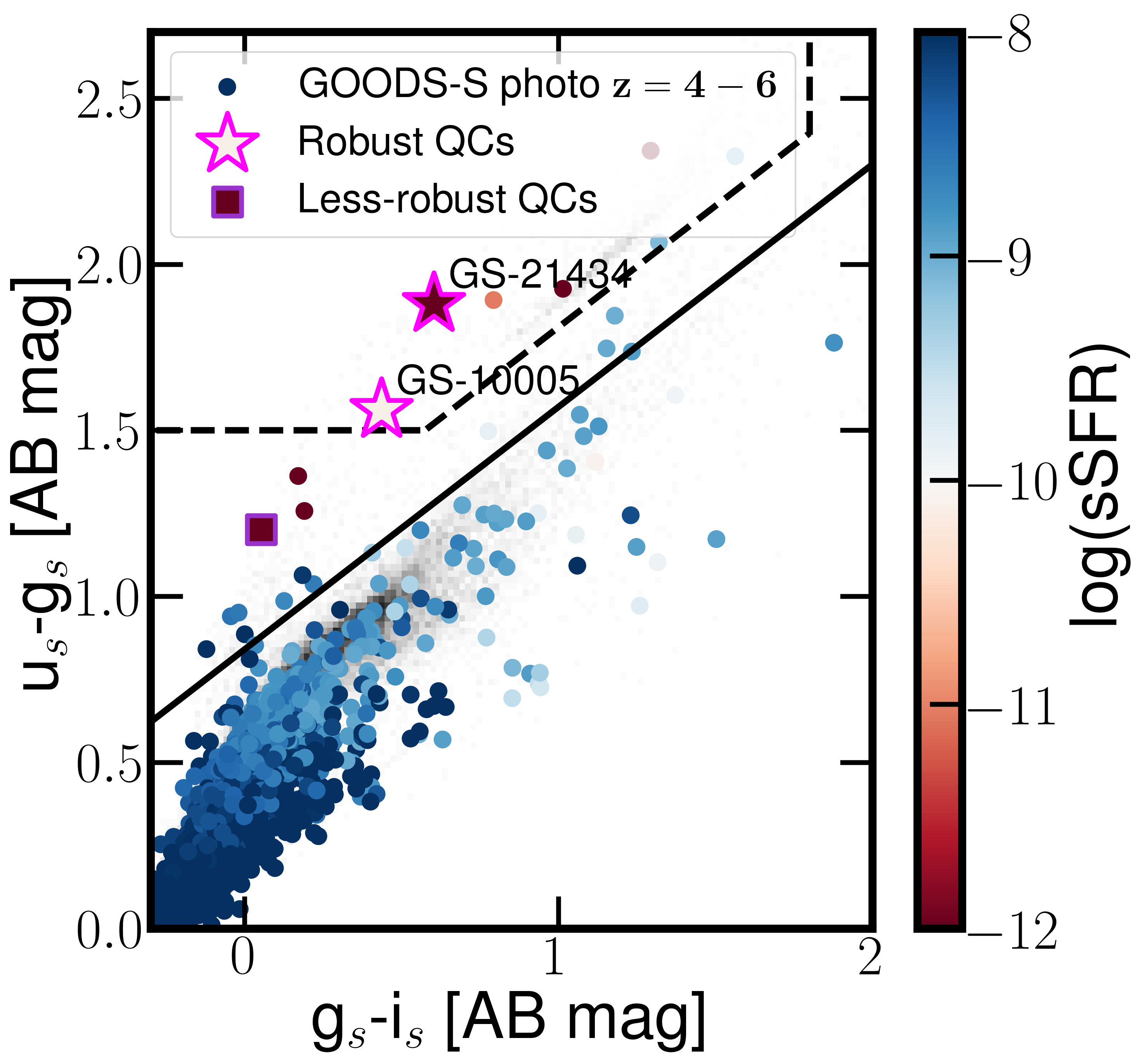}\\
    
    \includegraphics[width=0.4\textwidth]{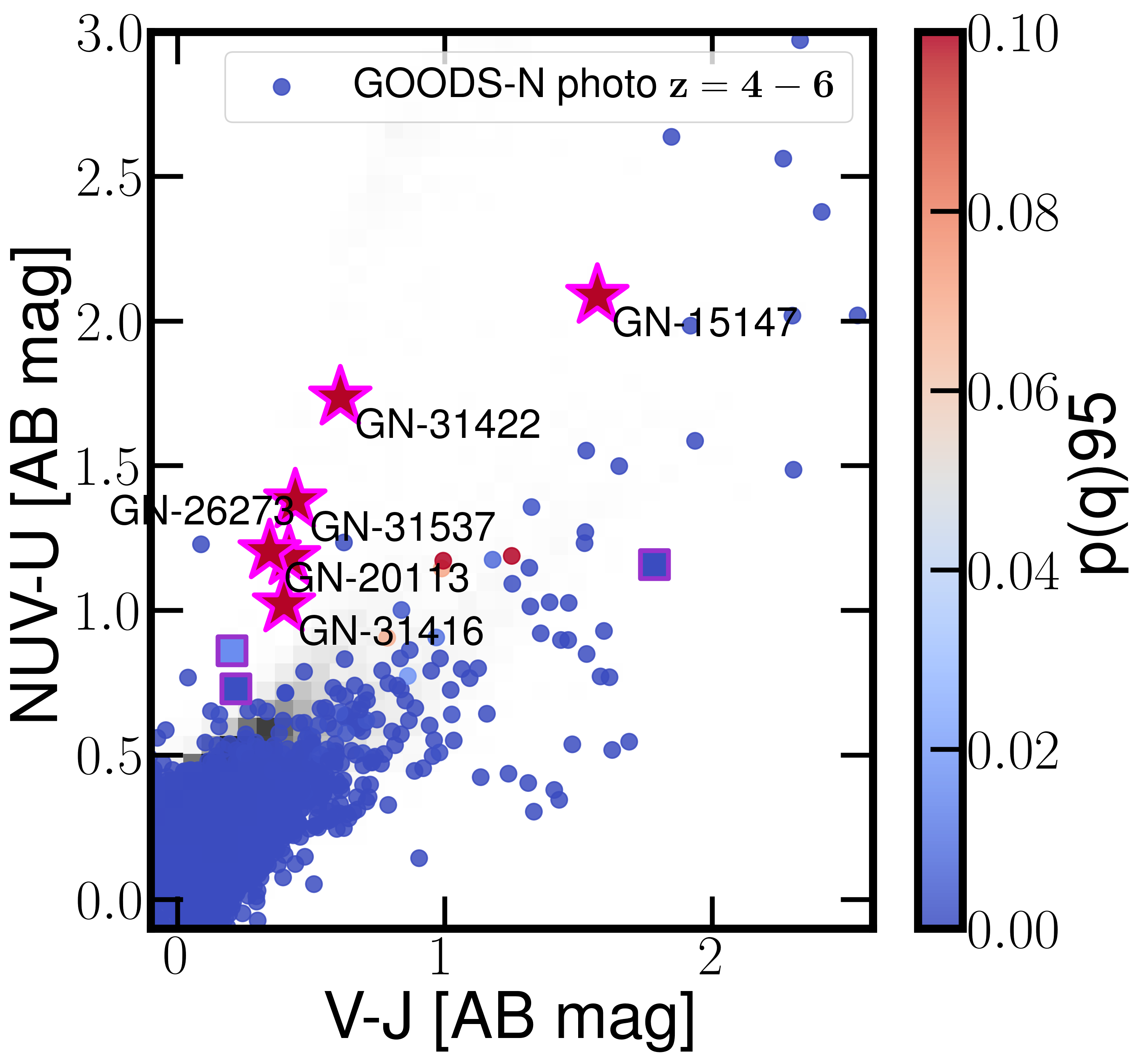}
    \hspace{0.01\textwidth}
    \includegraphics[width=0.4\textwidth]{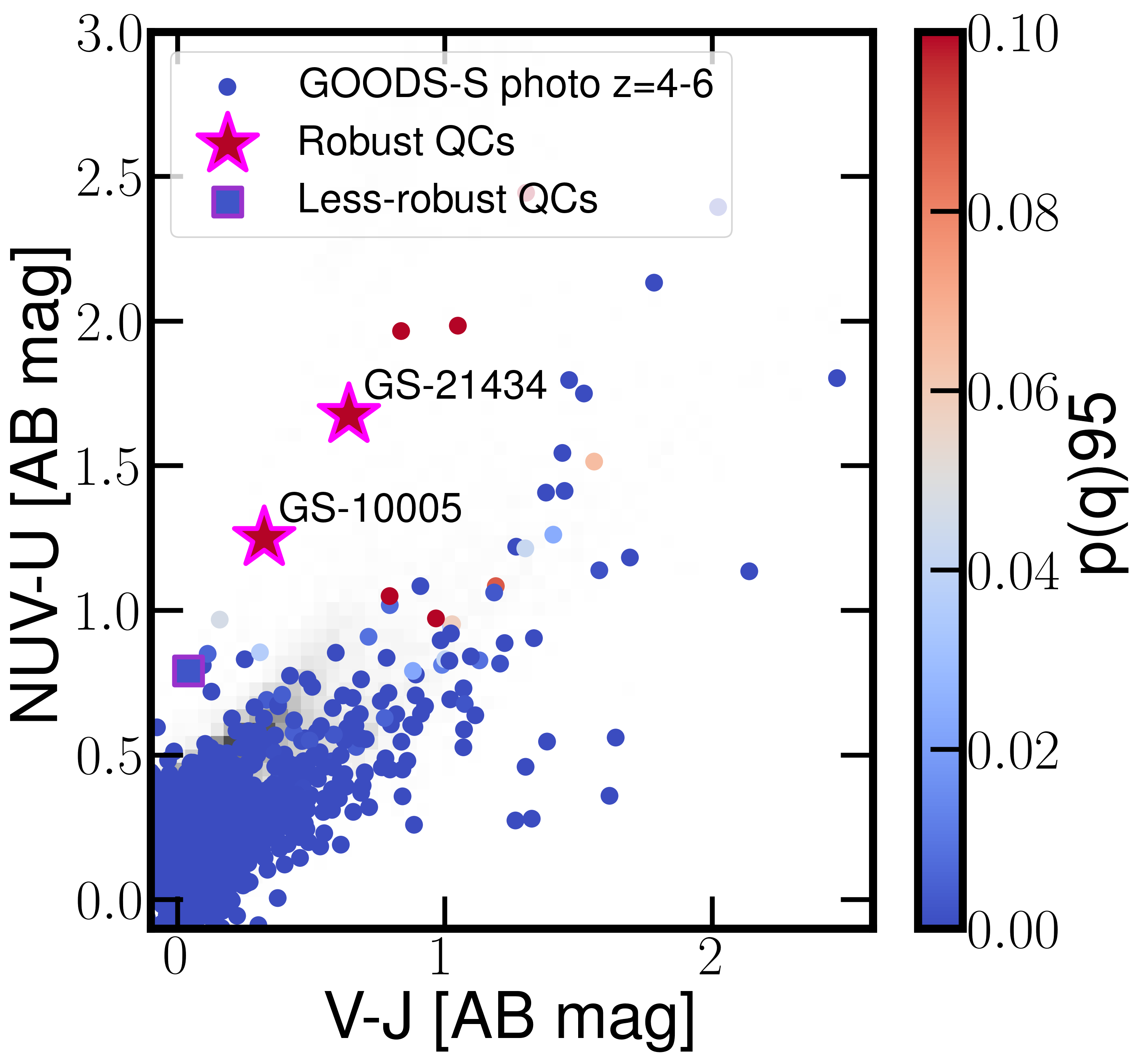}\\
    \caption{Color–color diagrams used to identify quiescent galaxy candidates in the GOODS-N (left column) and GOODS-S (right column) proto-clusters. The top panels show the rest-frame UVJ color--color selection. The quiescent region (black dashed line) is adopted from \citet{Belli2019}. We modify the selection to Equation \ref{eq:UVJ3} (black solid line) to include young, rare high-redshift quiescent galaxies. The middle panels show the synthetic $(ugi)_s$ color--color selection. We adapt the original quiescent box (black dashed line) from \citet{Antwi-Danso2023} to Equation \ref{eq:ugi2} (diagonal solid black line) to include younger quiescent galaxies. Lastly, the bottom panels correspond to the GMM/NUVU$-$VJ selection from \citet{Gould2023}. Robust quiescent candidates within the proto-clusters are marked as magenta-outlined stars, and the less-robust quiescent candidates in the proto-clusters are squares outlined in purple. The comparison samples from GOODS-N (left) and GOODS-S (right) at $z=4-6$ are color-coded by their specific star-formation rate in the top and middle panels and quiescent probability p(q) in the bottom panels. Gray background histograms show the distribution of all galaxies in our photometric catalogs. See Sections \ref{sec:uvj}, \ref{sec:ugi}, and \ref{sec:nuvu-vj} for more details about each selection method.}
    \label{fig:color-color}
\end{figure*}

\subsubsection{UVJ} \label{sec:uvj}
First, we select quiescent galaxies based on their rest-frame $U-V$ and $V-J$ colors, using the extended \citet{Schreiber2015} UVJ criterion \citep{Belli2019}. \citet{Belli2019} extended this selection along the fast-quenching track to include younger and post-starburst galaxies that reside below the horizontal edge of the \textit{UVJ} diagram \citep{Belli2019, Carnall2019}. This criterion helps mitigate biases against unobscured, recently quenched galaxies that retain significant UV emission from young, massive stars that have not yet evolved off the main sequence \citep{Marsan2015, Merlin2018, Merlin2019, Schreiber2018, Carnall2020}.

We show the extended UVJ cut from \citet{Belli2019} as the black dashed line defined by Equation \ref{eq:UVJ_} - \ref{eq:UVJ} in the top panels of Figure \ref{fig:color-color}. 

\begin{align}
    (U-V) &> 0.88\,(V - J) + 0.49 \label{eq:UVJ_} \\ 
    (V - J) &< 1.6 \label{eq:UVJ}
\end{align}

To capture the younger, recently quenched systems expected at early cosmic epochs (z$>$4) where the universe is only slightly more than one billion years old, we modify the canonical rest-frame color selection thresholds. Standard selection boundaries, such as the \citet{Schreiber2015, Belli2019} criteria (shown as the gray and black dashed lines, respectively, in Figure \ref{fig:color-color}), are optimized for mature, older quiescent populations dominant at z $<$ 2. At z$\sim 4-6$, however, the restricted age of the universe implies that quenched galaxies are intrinsically young and possess bluer rest-frame colors that place them directly below the traditional diagonal delimiter. Rather than a static lower-redshift boundary, we adopt an empirically motivated selection limit to ensure high selection completeness for these genuine, early quiescent systems. This modified \textit{UVJ} criterion is defined by (Equation \ref{eq:UVJ3}):

\begin{equation}
        (U - V )> 0.88\,(V - J ) + 0.25 
        \label{eq:UVJ3}
\end{equation}

This boundary is represented by the solid diagonal line in the top panels of Figure \ref{fig:color-color}. All of the final quiescent candidates in our sample are selected using this threshold.

Our approach aligns closely with recent JWST spectroscopic results from \citet{Baker2025}, who also identified a population of robustly quenched galaxies (confirmed via low sSFR and NIRSpec/PRISM spectra) residing outside the standard UVJ box. To accommodate these sources, \citet{Baker2025} incorporated 1$\sigma$ photometric uncertainties by perturbing the U, V, and J colors of their sample, ultimately defining an extended UVJ framework following \citet{Belli2019}. To encapsulate the entirety of their quiescent sample, their required empirical limit shifts to:
\begin{equation}
(U - V) > 1.30\,(V - J) - 0.20
\end{equation}

This criterion from \citet{Baker2025} is plotted as the gray dot-dashed line in the top panels of Figure \ref{fig:color-color}. Driven by the same physical goal of identifying young, rare high-redshift quiescent galaxies, our modified UVJ cut is remarkably similar to the boundary mapped out by \citet{Baker2025}. In fact, our selection successfully recovers all of the spectroscopically confirmed quiescent galaxies from their sample, with the sole exception of their single bluest and youngest source, which sits just outside our threshold. This source lies only $\sim$0.1 mag below our boundary in U-V, and is consistent with our selection within its 1$\sigma$ rest-frame color uncertainties. Accounting for these uncertainties, as \citet{Baker2025} did in defining their own boundary, our modified UVJ selection would recover their entire quiescent sample.

\subsubsection{Synthetic $(ugi)_s$} \label{sec:ugi}
We also adapt the novel $u_s-g_s$ and $g_s-i_s$, i.e., $(ugi)_s$, rest-frame color-color selection presented in \citet{Antwi-Danso2023}. The $(ugi)_s$ method has proven successful at selecting high-redshift quiescent galaxies in the JWST era by reducing contamination in quiescent samples at $z\sim3-6$ compared to the \textit{UVJ} selection. The method separates quiescent and star-forming galaxies by a box defined by Equations \ref{eq:ugi}: 
\begin{equation}
    \begin{split}
        (u_s - g_s)&>1.5, (g_s - i_s)<1.8\\
        (u_s - g_s)&>0.73\,(g_s - i_s) + 1.08 \\
        \label{eq:ugi}
    \end{split}
\end{equation}

We again modify the diagonal edge of the box to capture younger, post-starburst galaxies at high redshift, mirroring the shift we applied in the UVJ diagram, which gives the modified criterion in Equation \ref{eq:ugi2}:

\begin{equation}
        (u_s - g_s)> 0.88\,(g_s - i_s) + 0.84
        \label{eq:ugi2}
\end{equation}

The results of the $(ugi)_s$ color selection are shown in the middle panels of Figure \ref{fig:color-color}. All of our robust quiescent candidates, as well as 3 less-robust candidates, are selected by the $(ugi)_s$ method.

\subsubsection{GMM/NUVU$-$VJ} \label{sec:nuvu-vj}
We also utilize another rest-frame color selection method from \citet{Gould2023} using rest-frame $NUV-U$, $U-V$, $V-J$ colors which has been shown to outperform \textit{UVJ} selection for selecting quiescent galaxies at $z>3$. This selection process differs from the two previous rest-frame color selections (i.e., \textit{UVJ} and (ugi)$_s$) in two ways. First, the GMM/NUVU$-$VJ method uses four bands, producing three colors instead of two. The extra dimension provides additional information that helps probe light from recent star formation. Secondly, the separation of quiescent galaxies from star-forming is not binary and is not defined by a fixed “box”. Quiescent galaxies are selected based on their probability of belonging to the quiescent group defined by a Gaussian mixture model (GMM) trained on rest-frame colors ($NUV-U$, $U-V$, $V-J$) of massive galaxies at $2 < z < 3$. For further details about the methodology of GMM/NUVU$-$VJ quiescent selection and the training set, we refer the reader to \citet{Gould2023}.

Following this method, we calculate the quiescent probability p(q) for every galaxy that belongs to the four proto-clusters based on its photometric redshift. Of the 12 total quiescent candidates, 10 were selected using the GMM/NUVU$-$VJ method with p(q)$>$ 0.01. 

We also apply a more conservative threshold of p(q)$>$0.1, resulting in only 8 candidates meeting this stricter criterion. These higher-probability sources exhibit more prominent Balmer breaks and lower rest-frame UV emission. The results using GMM/NUVU$-$VJ method are shown in the bottom panels of Figure \ref{fig:color-color}.

\subsubsection{sSFR cut}  \label{sec:low-ssfr}
We impose a cut based on recent star formation activity over the past 100 Myr, as inferred from SED fitting, in addition to the rest-frame color–color selections. Specifically, we require
\begin{equation} \label{eq:ssfr}
    \mathrm{sSFR}_{100 Myr} \leq \frac{0.2}{t_{obs}}
\end{equation}
A galaxy is included in our quiescent sample if its sSFR satisfies this criterion within its 1$\sigma$ uncertainty, i.e., if the 16th percentile of its sSFR posterior lies below 0.2/t$_{obs}$.

\begin{figure}
    \hspace{-2em}
    \centering
    \includegraphics[width=0.9\linewidth]{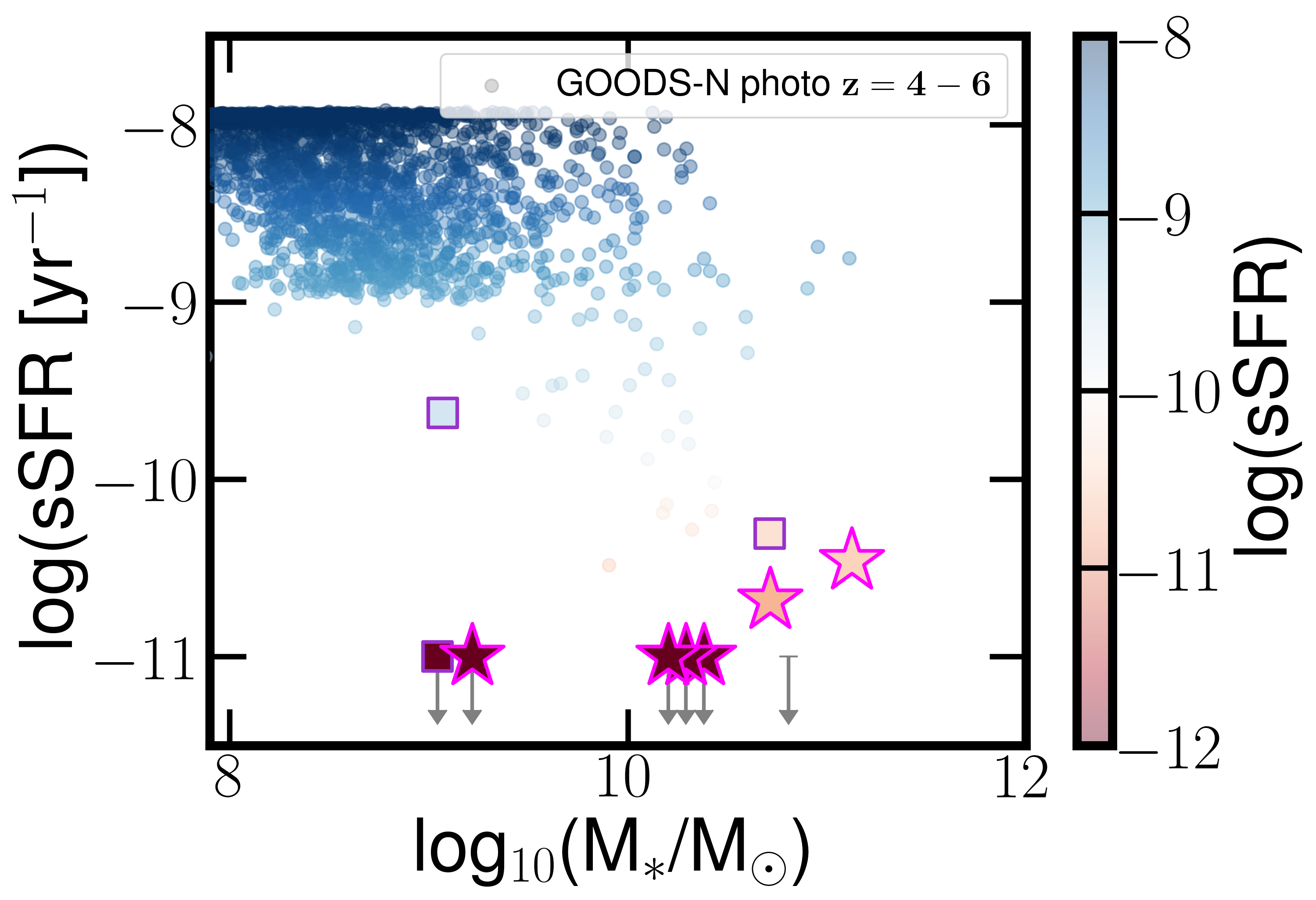}\\[2.5ex]
    \vspace{-1.0em}
    \hspace{-2em}
    \includegraphics[width=0.9\linewidth]{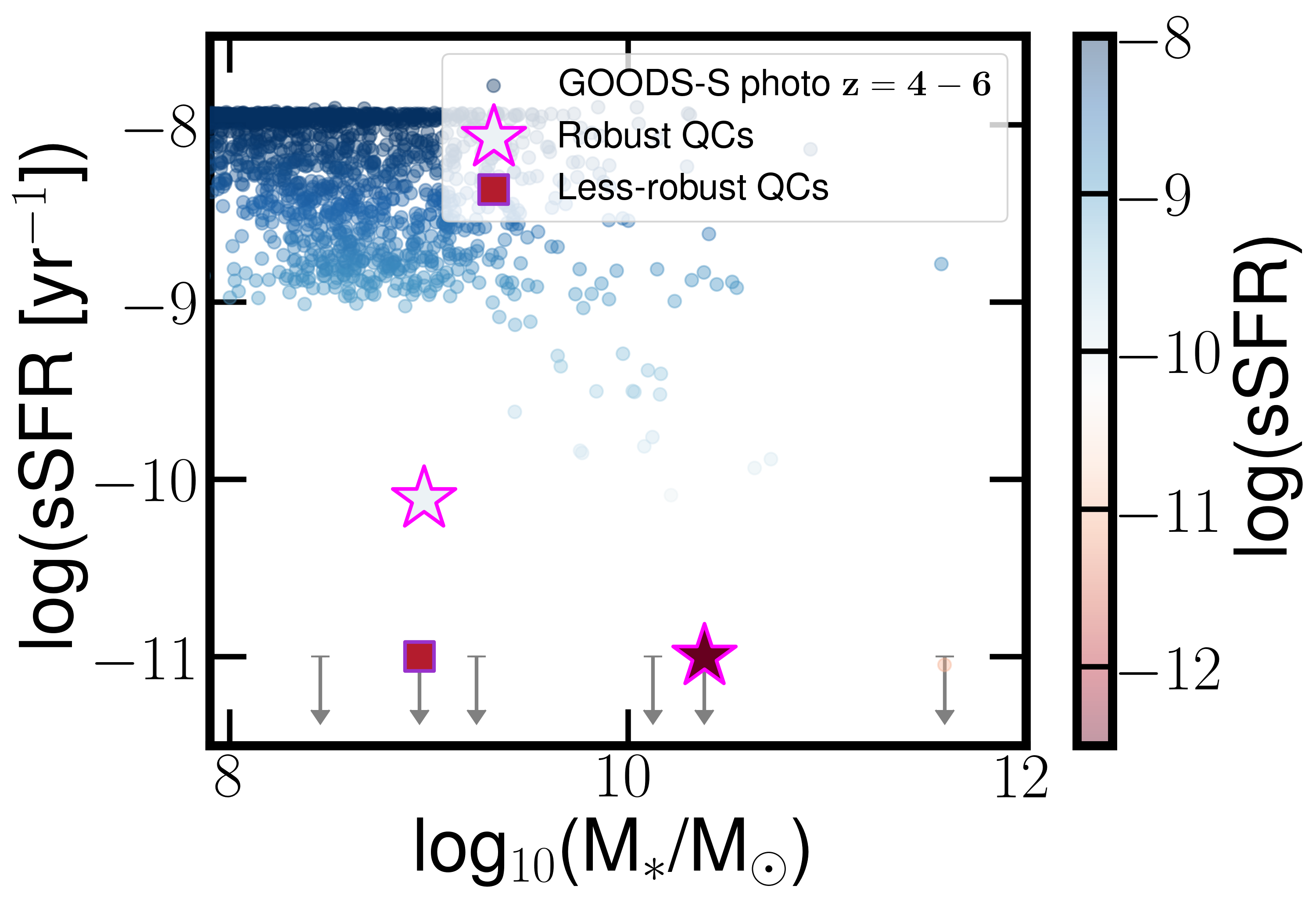}
    \caption{Specific star formation rate $\log$(sSFR) vs. stellar mass $\log$($M_*$). The robust quiescent candidates within the proto-clusters are shown as stars, color-coded by their specific star-formation rate. Other quiescent candidates in the proto-clusters are shown in squares, color-coded in  specific star-formation rates. The comparison consists of galaxies at $z=4-6$ in GOODS-N (top) and GOODS-S (bottom). If a galaxy has $\log$(sSFR) $<-11$, we mark it as an upper limit at $\log$(sSFR) $= -11$.}
    \label{fig:ssfr-mass}
\end{figure}

This selection (Equation~\ref{eq:ssfr}) has been widely used in recent studies to identify quiescent galaxies at high redshift \citep{Carnall2024, Baker2025}, and is broadly equivalent to the rest-frame UVJ color selection \citep{Gallazzi2014, Carnall2018, Carnall2020}. We show the sSFR of our final quiescent sample as a function of stellar mass in Figure \ref{fig:ssfr-mass}.

After the sources are selected by the above methods, we visually inspect their RGB images, SEDs, and SFHs. We expect distant quiescent galaxies to appear as smooth, compact red or red-orange sources. In the SEDs, we look for the lack of UV emission and rest-frame optical emission lines, which indicates no recent star-formation activity, and their corresponding SFHs should show little to no SFR at the time of observation. Lastly, we confirm that there is little or no significant H$\alpha$ emission. We reject any source that does not satisfy all of the visual checks to finalize the sample of quiescent galaxies. For example, several GOODS-S galaxies that fall into the UVJ quiescent box and have very low sSFR are rejected due to a poor SED fit, the presence of H$\alpha$ emission, or its SED resembles that of a dusty-SF galaxy (See the top right panel of Figure \ref{fig:color-color}). 

\figsetstart
\figsetnum{4}
\figsettitle{Robust Quiescent Candidates}

\figsetgrpstart
\figsetgrpnum{4.1}
\figsetgrptitle{GN-31537}
\figsetplot{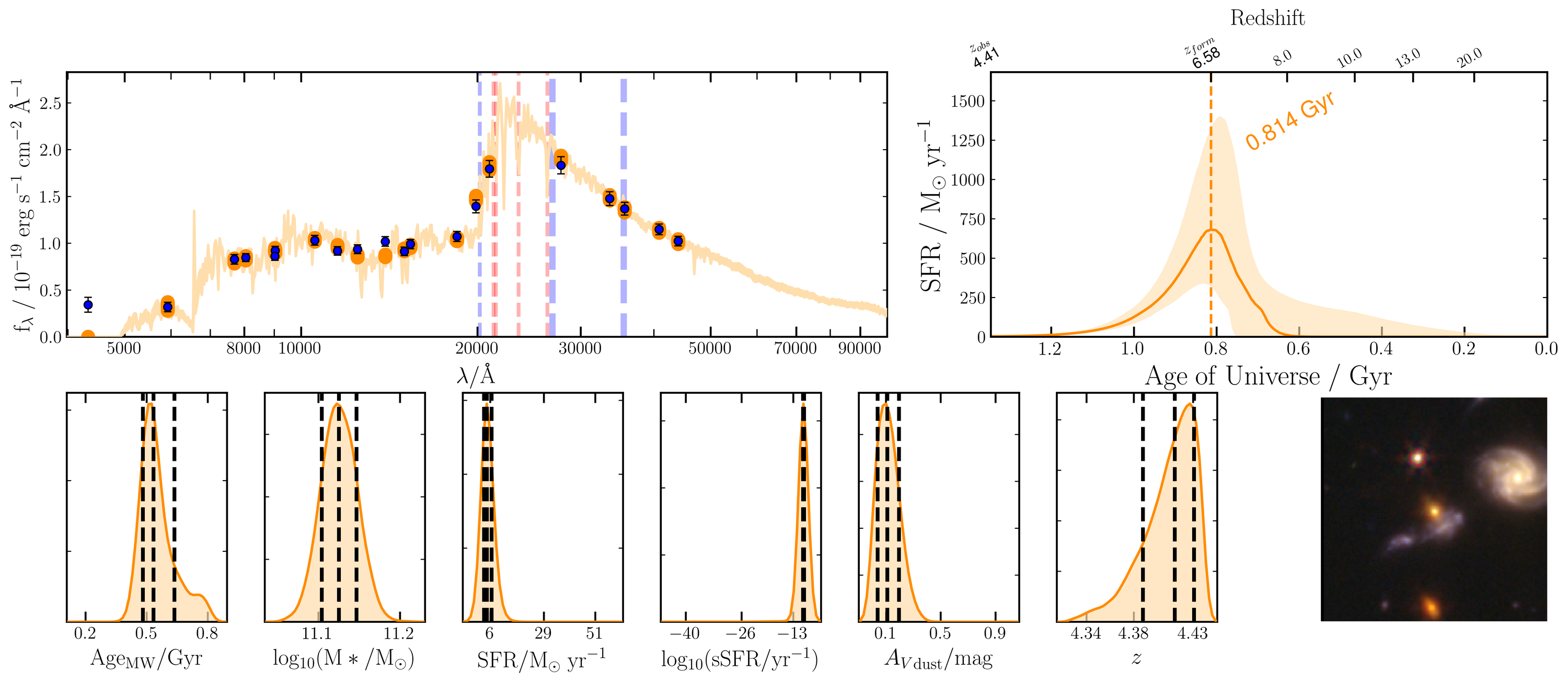}
\figsetgrpend

\figsetgrpstart
\figsetgrpnum{4.2}
\figsetgrptitle{GN-15147}
\figsetplot{GN-15147_master.png}
\figsetgrpend

\figsetgrpstart
\figsetgrpnum{4.3}
\figsetgrptitle{GN-31422}
\figsetplot{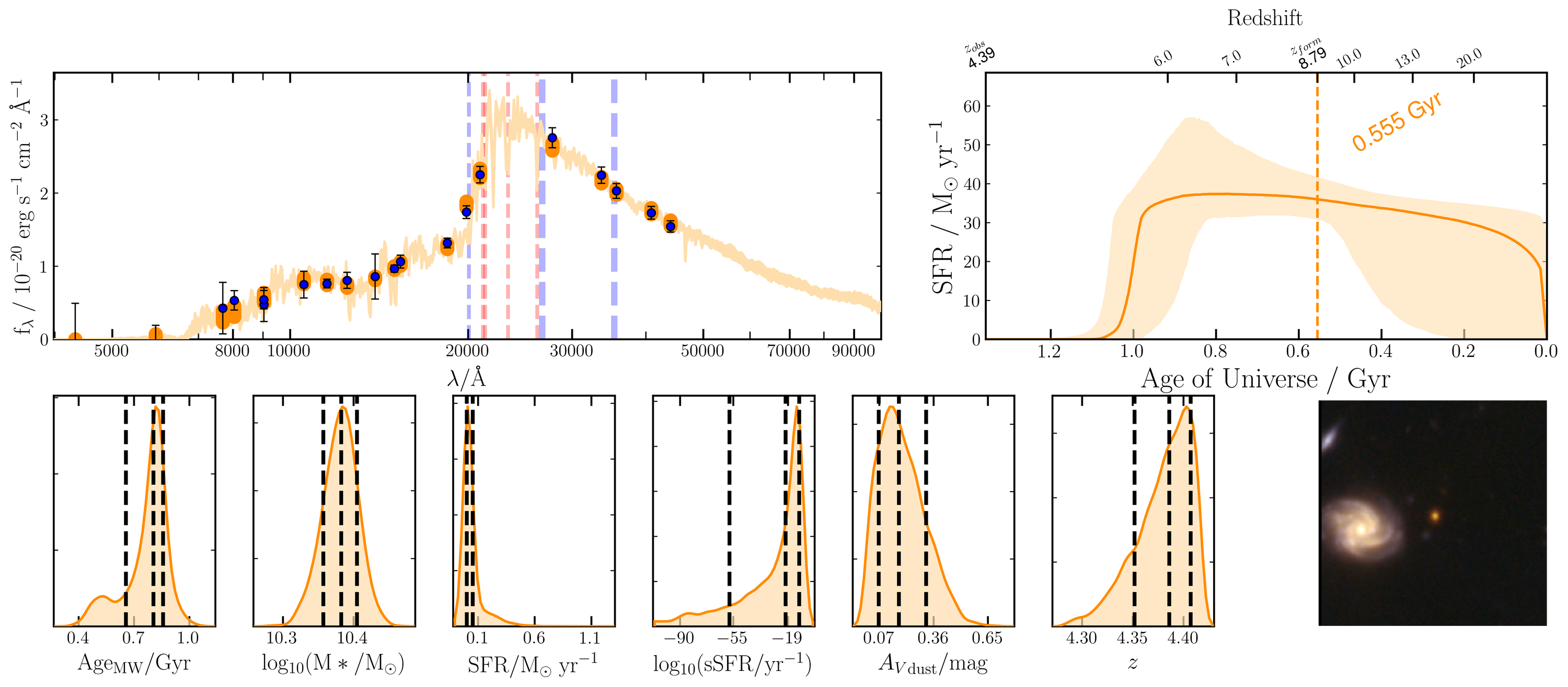}
\figsetgrpend

\figsetgrpstart
\figsetgrpnum{4.4}
\figsetgrptitle{GN-20113}
\figsetplot{GN-20113_master.png}
\figsetgrpend

\figsetgrpstart
\figsetgrpnum{4.5}
\figsetgrptitle{GN-31416}
\figsetplot{GN-31416_master.png}
\figsetgrpend

\figsetgrpstart
\figsetgrpnum{4.6}
\figsetgrptitle{GS-10005}
\figsetplot{GS-10005_master.png}
\figsetgrpend
\figsetend

\begin{figure*}
        \centering
        \vspace{-1em}
        \textbf{GN-31537}\\
        \vspace{-1em}
        \includegraphics[width=0.95\linewidth]{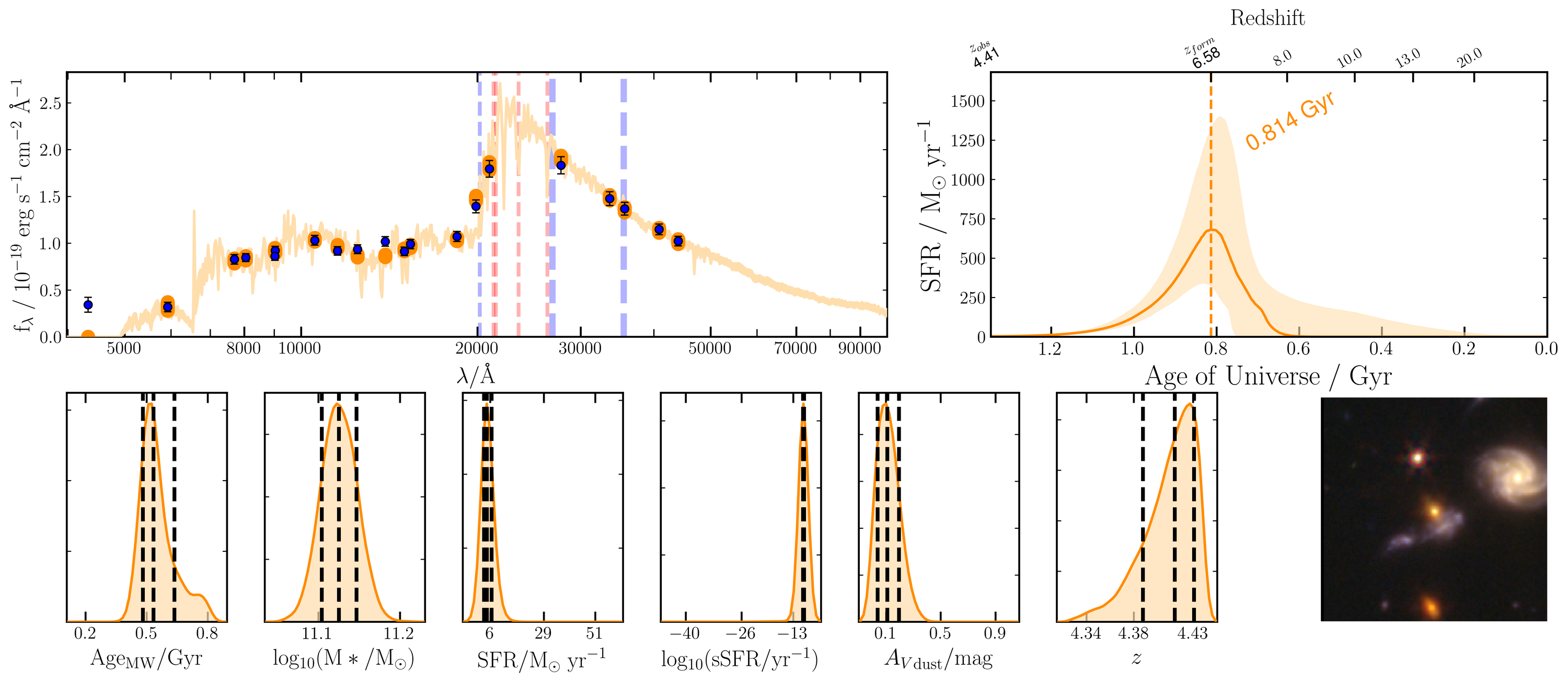}

        \textbf{GN-31422}\\
        \vspace{-1em}
        \includegraphics[width=0.95\linewidth]{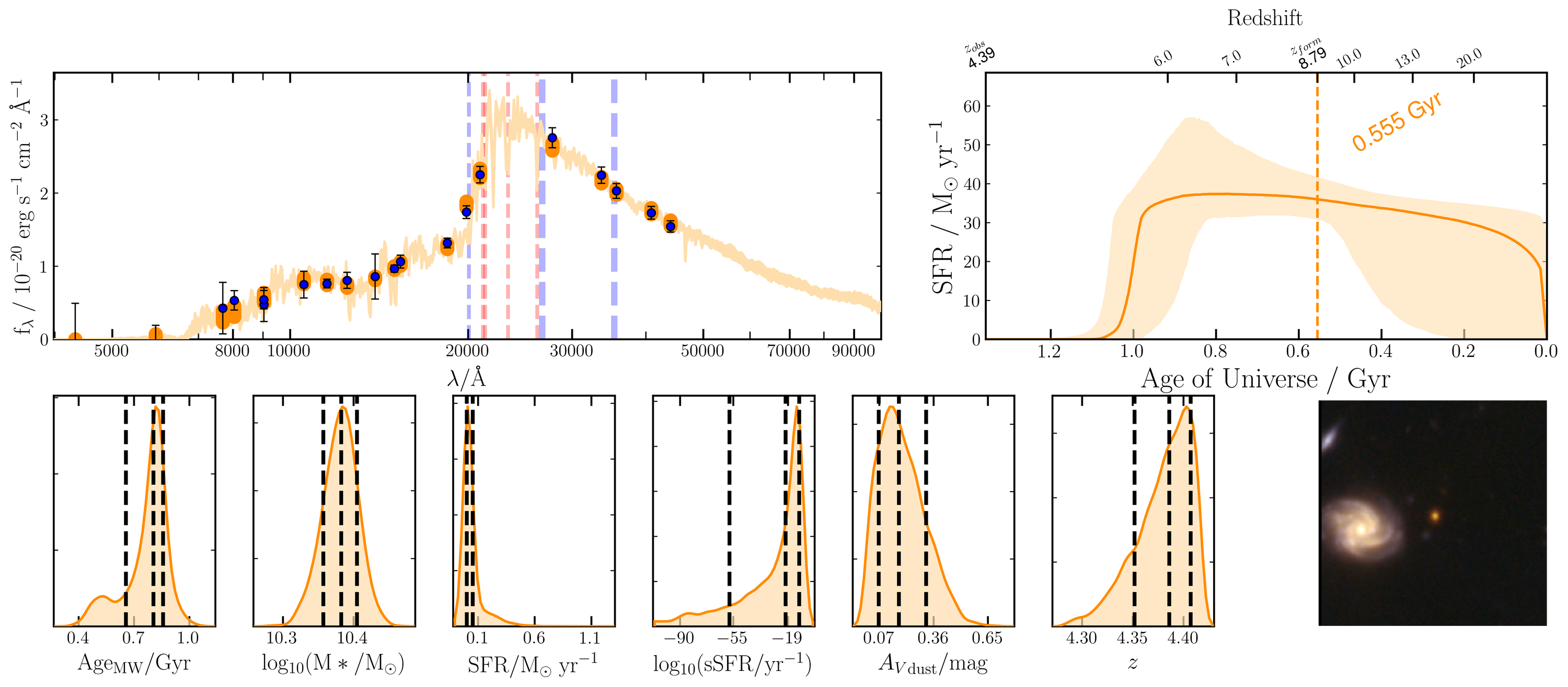}
        
        \caption{Examples of robust quiescent galaxies. The complete figure set (6 images), containing all robust candidates with photometric redshifts, is available in the online journal. Observed photometry is shown in blue, and the posterior SEDs from BAGPIPES are shown in orange. The vertical blue dashed lines mark the wavelengths of emission features, including [OII] and [OIII] doublets, and H$\alpha$ + [NII]. The red dashed lines mark the wavelengths of absorption features, including Ca H+K, H$\beta$, and H$\gamma$. The top-right panels show the star-formation histories from the Big Bang (right) to the time when the galaxy is observed (left). The vertical orange dashed line marks $t_{form}$ in Gyr and its corresponding $z_{form}$, when the galaxy formed $50\%$ of its mass. The RGB cutouts are composed of the F182M, F210M, and F444W bands with a box size of 4" $\times$ 4". The bottom panels show the posterior distributions of mass-weighted age, stellar mass, SFR, sSFR, Av, and $z$ (from left to right), with the vertical dashed lines representing the 16th, 50th, and 84th percentiles.} \label{fig:combined-seds}
 
\end{figure*}

\section{Results} \label{sec:results}
\subsection{Quiescent Candidates}
We identify a total of \textbf{12} quiescent candidates across the four proto-clusters using the selection methods described above. Of these, 3 are associated with the proto-cluster at $z=4.267$ (GOODS-S), 5 at $z=4.406$ (GOODS-N), 4 at $z=5.194$ (GOODS-N), and none at $z=5.384$ (GOODS-S). Their stellar masses range from $\log$(M$_*/M_{\odot}) = 8.95$ to $\log$(M$_*/M_{\odot}) = 11.13$. 

Among the 12 quiescent candidates, 8 galaxies (namely, GN-31537, GN-15147, GN-31422, GN-26273, GN-31416, GN-20113, GS-21434, and GS-10005) are considered robust, satisfying all five selection criteria (M1-M5 in Table \ref{table:quiescent-candidates}). Notably, $75.0\%$ (6 out of 8) of our robust candidates are associated with the $z=4.267$ and $z=4.406$ proto-clusters. Example SEDs of robust quiescent galaxies are shown in Figure \ref{fig:combined-seds}. 

Table \ref{table:quiescent-candidates} summarizes our final set of quiescent candidates and the selection criteria they satisfy. We also report their redshift ($z$), stellar mass ($\log$(M$_*$)), star-formation rate ($\log$(SFR)), specific star-formation rate ($\log$(sSFR)), and mass-weighted age (Age$_{mw}$) derived from BAGPIPES SED modeling. 

We present all the SEDs, SFHs, and cutouts of our quiescent galaxies in Figure Set 9 (Appendix \ref{app:sed}). Coordinates of the quiescent galaxies are provided in Appendix Table \ref{table:coordinates}.

\subsection{Spatial Distribution in Proto-clusters}

\begin{figure*}
    \centering
    \includegraphics[width=.9\textwidth]{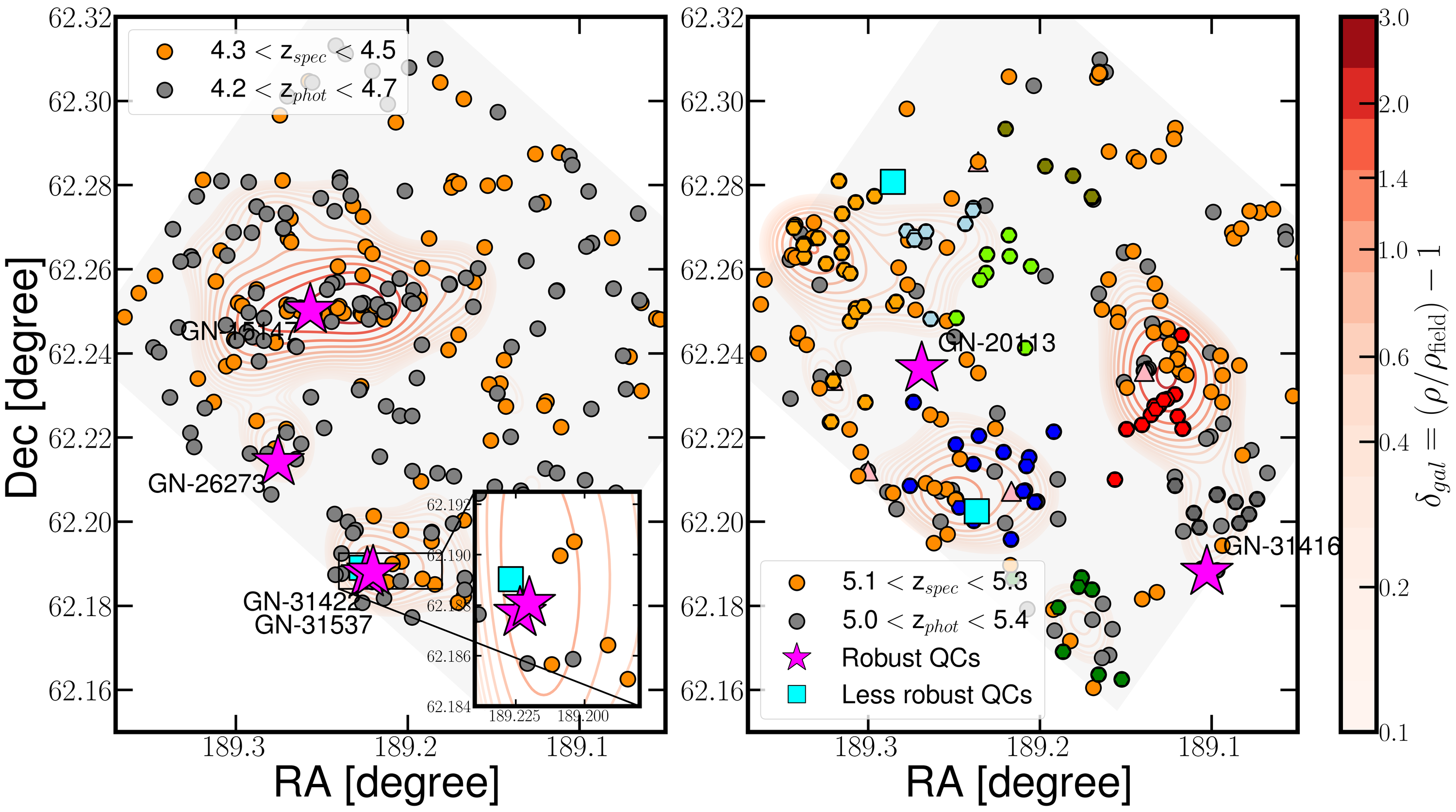}
    \caption{Location of proto-cluster members and quiescent galaxies within the GOODS-N proto-clusters at $z=4.406$ (left) and $z=5.194$ (right) in RA and Dec. The robust quiescent galaxies are shown as magenta stars. Other quiescent candidates are shown as cyan square. In the $z=4.406$ proto-cluster (left), the proto-cluster members include H$\alpha$ emitting galaxies with spectroscopic redshifts at $4.3 < z_{\rm spec} < 4.5$ \citep{Alba2014} in orange and galaxies with photometric redshifts $4.2 < z_{phot} < 4.7$ in gray. In the $z=5.194$ proto-cluster, the members include H$\alpha$-emitting galaxies with spectroscopic redshift $5.1 < z_{spec} < 5.3$ in orange and galaxies with photometric redshifts $5.0 < z_{\rm phot} < 5.4$ in gray. We also highlight the sub-structures defined in \citet{Herard-Demanche2023} in varying color dots (red, blue, green, dim gray, purple, light blue, lawn green, and olive). The overlaid contour lines and color bar represent the spatial galaxy number overdensity contrast, defined as: $\delta_{\rm gal} = (\rho /\rho_{\rm field}) - 1$, where $\rho$ is the local surface density computed via a Gaussian Kernel Density Estimate (KDE) and $\rho_{\rm field}$ is the average background number density across the FRESCO survey area. The color bar scales logarithmically and begins at  $\delta_{\rm gal} = 0.1$, representing regions 10\% overdense relative to the field. The high-contrast regions trace the peaks of the overdensities. The gray shaded region in the background marks the coverage of FRESCO in GOODS-N.}
    \label{fig:gdn-overdensity-all}
\end{figure*}

\begin{figure*}
    \centering
    \includegraphics[width=.9\textwidth]{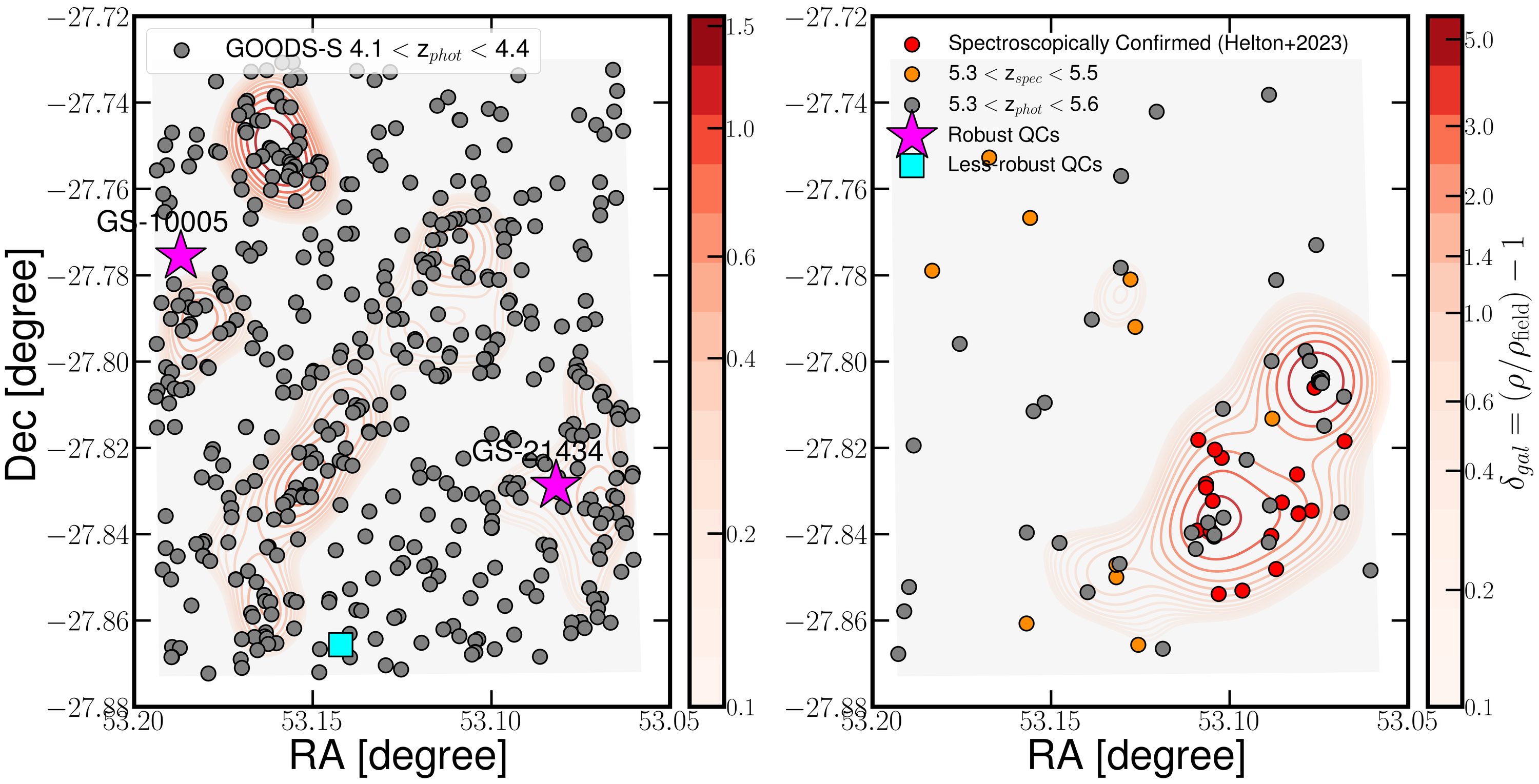}
    \caption{Location of proto-cluster members and quiescent galaxies within the GOODS-S proto-clusters at $z_{od, phot} = 4.267$ (left) and $z_{od} = 5.384$ (right) in RA and DEC. In the $z_{od, phot} \sim 4.267$ proto-cluster, the members are galaxies in our photometric catalog with photometric redshift in the range $4.1 < z_{\rm phot} < 4.4$ shown in gray. In the $z_{od} = 5.384$ proto-cluster, the  proto-cluster members include H$\alpha$ emitters with spectroscopic redshift  $5.3 <z_{\rm spec} < 5.5$ and galaxies with photometric redshift $5.3 <z_{\rm phot} < 5.6$. We also highlight spectroscopically confirmed overdensity members from \citet{Helton2024} in red. The contours show the overdensity of protocluster members. The gray shaded region in the background marks the coverage of FRESCO in GOODS-S. Symbols and contours are the same as in Figure \ref{fig:gdn-overdensity-all}.}
    \label{fig:gds-overdensity-all}
\end{figure*}

We map out the spatial distribution of our quiescent galaxies within the four proto-clusters in Figure \ref{fig:gdn-overdensity-all} and Figure \ref{fig:gds-overdensity-all}. We find that the proto-clusters are broken down into smaller substructures, which is expected at these redshifts. Similar features are also seen in previous simulations and observations (e.g., \citealt{Chiang2017, Herard-Demanche2023}). 

Interestingly, the quiescent galaxies often appear to lie near the edges of the overdensity peaks. At lower redshifts, we typically expect massive quiescent galaxies to reside near the centers of massive halos \citep{Collins2009}. However, at these higher redshifts, proto-clusters are not yet virialized, and their substructures require time to collapse and evolve into the massive galaxy clusters we see at $z\sim0$. This likely explains why the locations of the quiescent galaxies do not always coincide with the peaks of the overdensities. 

Additionally, these proto-clusters were identified using star-forming, line-emitting galaxies. This method, while effective, could introduce observational bias and may not accurately represent the full galaxy population or the underlying dark matter distribution. 

We also show the structures of proto-clusters weighted by galaxy stellar masses in appendix Figure \ref{fig:overdensity-all-mweighted}. We still see that the proto-clusters are fragmented into sub-components. In the proto-cluster at $z=4.406$ in GOODS-N, there are three quiescent galaxies (two of which are robust and massive) in very close proximity to each other, forming a peak in the overdensity. Also, a couple of small sub-components are centered around a single massive quiescent galaxy (i.e., at $z = 5.194$ in GOODS-N and $z\sim4.267$ in GOODS-S). This is not unexpected since the contours are mass-weighted and the quiescent galaxies are among the most massive galaxies in proto-clusters. Nevertheless, there is no consistent trend for quiescent galaxies to reside at the peak of overdensities, as many remain scattered toward the outskirts.

\subsection{Quiescent Fraction}
We calculate the quiescent fractions in the four proto-clusters in GOODS-N and GOODS-S, i.e., the ratio of the number of quiescent candidates to the total number of galaxies in each proto-cluster, 

\begin{equation}
        f_{\mathrm{QG}} = \frac{N_{\mathrm{Q}}}{N_{\mathrm{tot}}}
        \label{eq:QF}
\end{equation}

In Table \ref{table:quiescent_fraction}, we present the quiescent fractions, including the numerators and denominators calculated using Equation \ref{eq:QF} across three stellar mass intervals: $\log(M_*/{\rm M}_\odot) = 9.0\text{--}9.75$, $9.75\text{--}10.5$, and $10.5\text{--}11.25$. Two candidates in the z $= 4.267$ proto-cluster, GS-10005 ($\log(M_*/{\rm M}_\odot) = 8.98^{+0.04}_{-0.04}$) and GS-29940 ($\log(M_*/{\rm M}_\odot) = 8.95^{+0.04}_{-0.05}$), lie just below the lower edge of our lowest mass bin. We include them in the $9.00-9.75$ bin, as their stellar masses are consistent with the bin edge within their 1$\sigma$ uncertainties (to within 0.01 dex). The number of quiescent galaxies, $N_{Q}$, represents the galaxies in our final quiescent sample selected via the method described in Section \ref{sec:quiescent-selection} for each respective proto-cluster and $0.75\,\text{dex}$ stellar mass bin. The total number of galaxies, $N_{\text{tot}}$, includes all galaxies that have the flag \texttt{use\_phot=1}, $\text{S/N} > 3$ in the F444W band, and photometric redshifts falling within each proto-cluster range defined in Table \ref{table:overdensity-zphot}. Our stellar mass completeness limits range from $\log(M_{*, \rm lim}/{\rm M}_\odot) = 7.5$ at $z \sim 4$ to $\log(M_{*, \rm lim}/{\rm M}_\odot) = 9.0$ at $z \sim 9$ \citep{Weibel2024}.

With our 12 quiescent candidates, we calculate average quiescent fractions of $1.0\%^{+0.7}_{-0.4}$ (5/502), $3.5\%^{+2.8}_{-1.7}$ (4/115), and $33.3\%^{+22.2}_{-17.5}$ (3/9) in stellar mass bins of $\log$(M$_{*}$/M$_{\odot}$) = 9 -- 9.75, 9.75 -- 10.5, 10.5 -- 11.25, respectively, across the four proto-clusters. We compute the confidence intervals on the quiescent fractions using the small-number statistics of \citet{Gehrels1986}. For mass bins with fewer than 10 galaxies (N$_{tot} <$ 10), we estimate the error bar using Table 6 in \citet{Gehrels1986} for binomial statistics. 

\begin{figure*}[t]
    \centering
    \includegraphics[width=0.95\linewidth]{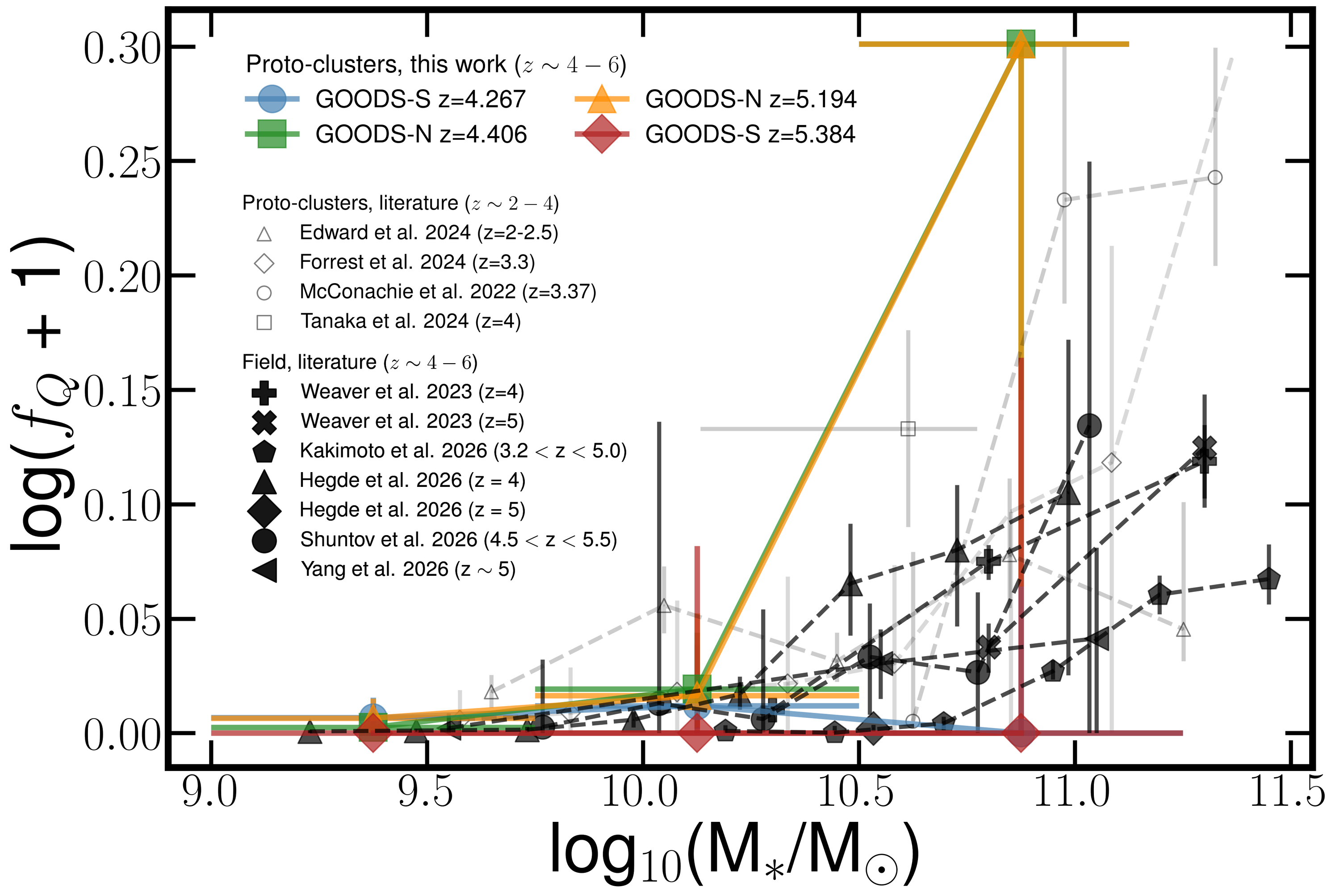}
    \caption{Quiescent fractions, $f_{Q}$, of the four proto-clusters are shown on a $\log_{10}$($f_{Q}$ + 1) scale, across three stellar mass bins $\log$(M$_{*}$/M$_{\odot}$) = 9.00 -- 9.75, 9.75 -- 10.50, and 10.50 -- 11.25. Note that a quiescent fraction of 100\% corresponds to $\sim 0.3$ on this scale of $\log$$_{10}$($f_{Q} +1)$. The quiescent fractions $\log_{10}$($f_{Q} +1)$ of our proto-clusters are shown as colored symbols connected by solid lines, color-coded by increasing proto-cluster redshift:  $z\sim4.267$ (GOODS-S) blue, $z=4.406$ (GOODS-N) green, $z=5.194$ (GOODS-N) orange, and $z=5.384$ (GOODS-S) red. Error bars are computed using the approximate confidence intervals for small-number statistics from \citet{Gehrels1986}. For statistically small mass bins (N$_{tot} <$ 10), we estimate the error bar using Table 6 in \citet{Gehrels1986} for binomial statistics. We compare our results with quiescent fractions reported in previous studies of proto-clusters at $z\sim$2--4 \citep{Forrest2024a,Edward2024,McConachie2022,Tanaka2024}, shown as open gray symbols. We also present quiescent fractions from \citet{Weaver2023, Kakimoto2026, Hegde2026, Yang2026, Shuntov2026} as the comparison set for non-overdense regions (i.e., field sample), shown as filled black symbols connected by dashed lines.}
    \label{fig:quiescent-fraction}
\end{figure*}

As shown in Figure \ref{fig:quiescent-fraction}, the quiescent fraction broadly increases with stellar mass. However, at the high-mass end, there is significant variation in the quiescent fraction from one proto-cluster to another. Specifically, the proto-clusters in GOODS-N at $z=4.406$ and $z=5.194$ exhibit very high quiescent fractions. In contrast, at similar redshifts in GOODS-S, the apparent absence of high-mass quiescent galaxies highlights the considerable proto-cluster to proto-cluster variation. At lower masses $\log$(M$_{*}$/M$_{\odot}$) $< 10.5$, the quiescent fractions are comparable across all four overdensities.

We also compare our quiescent fraction with previous studies of lower-redshift proto-clusters \citep{Forrest2024a, Edward2024, McConachie2022, Tanaka2024}. As seen in Figure \ref{fig:quiescent-fraction}, the large variation among proto-clusters at the high mass  $\log$(M$_{*}$/M$_{\odot}$) $> 10.5$ end is also apparent at $z\sim2-4$, whereas, the quiescent fractions at low masses remain comparable.

Additionally, we compare our quiescent fractions in proto-clusters to those of field galaxies at similar redshifts, z$\sim 4-6$. At the low-mass end, the quiescent fractions of field galaxies and proto-cluster galaxies are similar. They diverge at the high-mass end, $\log$(M$_{*}$/M$_{\odot}$) $> 10.5$, where the GOODS-N proto-clusters show elevated quiescent fractions compared to the field \citep{Weaver2023, Kakimoto2026, Hegde2026, Yang2026, Shuntov2026}.

\begin{deluxetable}{ccccc}
    \tabletypesize{\scriptsize} 
    \tablecaption{Quiescent fractions in the four proto-clusters\label{table:quiescent_fraction}}
    \tablehead{
    \colhead{} & \colhead{GOODS-S} & \colhead{GOODS-N} & \colhead{GOODS-N} & \colhead{GOODS-S} \\
    \colhead{$\log$($M_*$)} & \colhead{$z_{od} =4.267$} & \colhead{$z_{od}=4.406$} & \colhead{$z_{od}=5.194$} & \colhead{$z_{od}=5.384$}
    }
    \startdata
    $9.00 - 9.75$    & $0.016_{-0.010}^{+0.021}$ (2/128) & $0.006_{-0.005}^{+0.013}$ (1/175) & $0.015_{-0.010}^{+0.020}$ (2/132)&  $0.0_{-0.0}^{+0.028}$ (0/67) \\
    $9.75 - 10.50$   & $0.028_{-0.023}^{+0.065}$ (1/36) & $0.045_{-0.030}^{+0.061}$ (2/44) & $0.038_{-0.033}^{+0.090}$ (1/26) & $0.0_{-0.0}^{+0.207}$ (0/9) \\
    $10.50 - 11.25$  & $0.0_{-0.0}^{+0.459}$ (0/3)       & $1.0_{-0.602}^{+0.0}$ (2/2)     & $1.0_{-0.841}^{+0.0}$ (1/1)   & $0.0_{-0.0}^{+0.459}$ (0/3) \\
    \enddata
    \tablecomments{Quiescent fractions are shown across three stellar mass bins and ordered by increasing redshift from left to right: $z = 4.267$ (GOODS-S), $z=4.406$ (GOODS-N), $z=5.194$ (GOODS-N), and $z=5.384$ (GOODS-S). The integers that go into the numerators and denominators are quoted in parentheses. Error bars represent small-number Poisson uncertainties following \citet{Gehrels1986}.}
\end{deluxetable}

\subsection{Spectroscopic Confirmation with DJA}\label{sec:dja}
\begin{figure*}[t!]
    \centering

    \textbf{GS-21434}\\
    \vspace{-2em}
    \includegraphics[width=\linewidth]{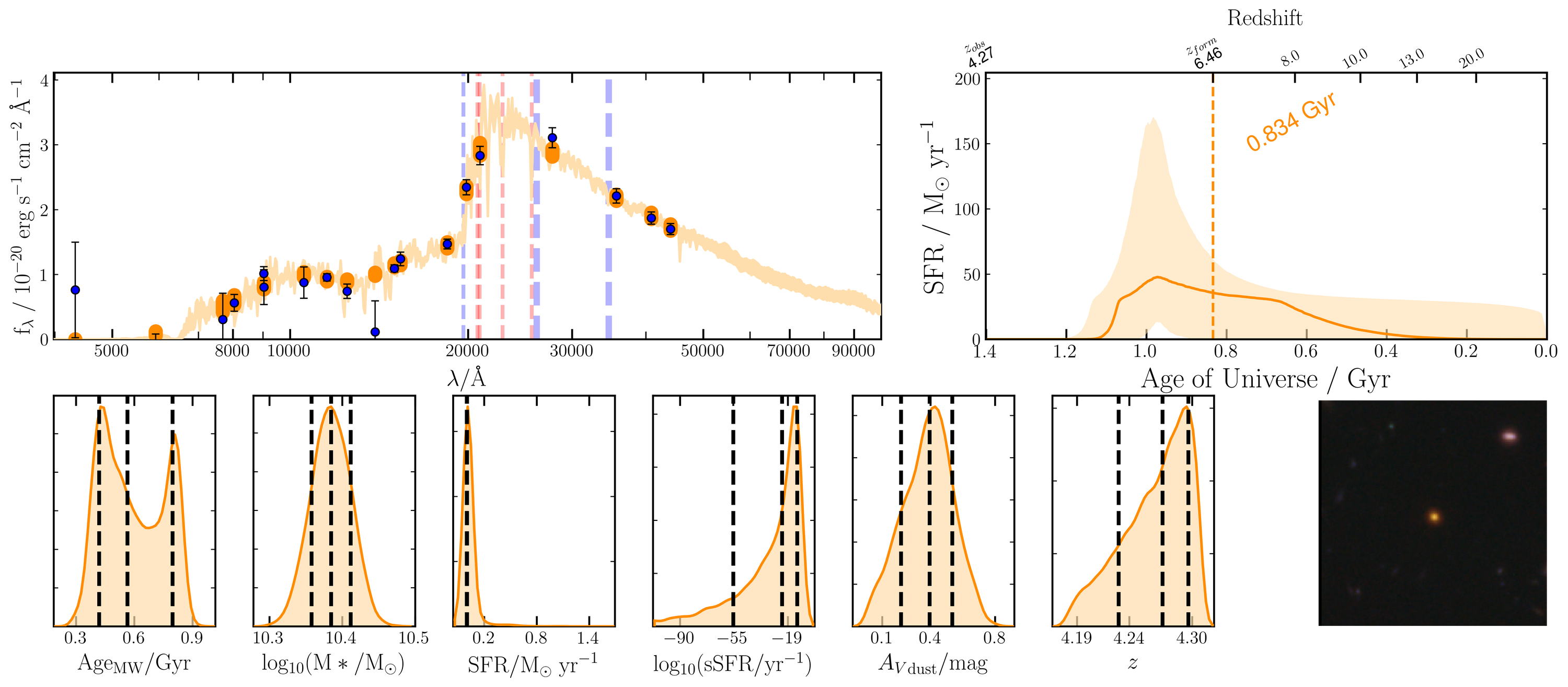}\\[1.5ex]

    \textbf{GN-26273}\\
    \vspace{-2em}
    \includegraphics[width=\linewidth]{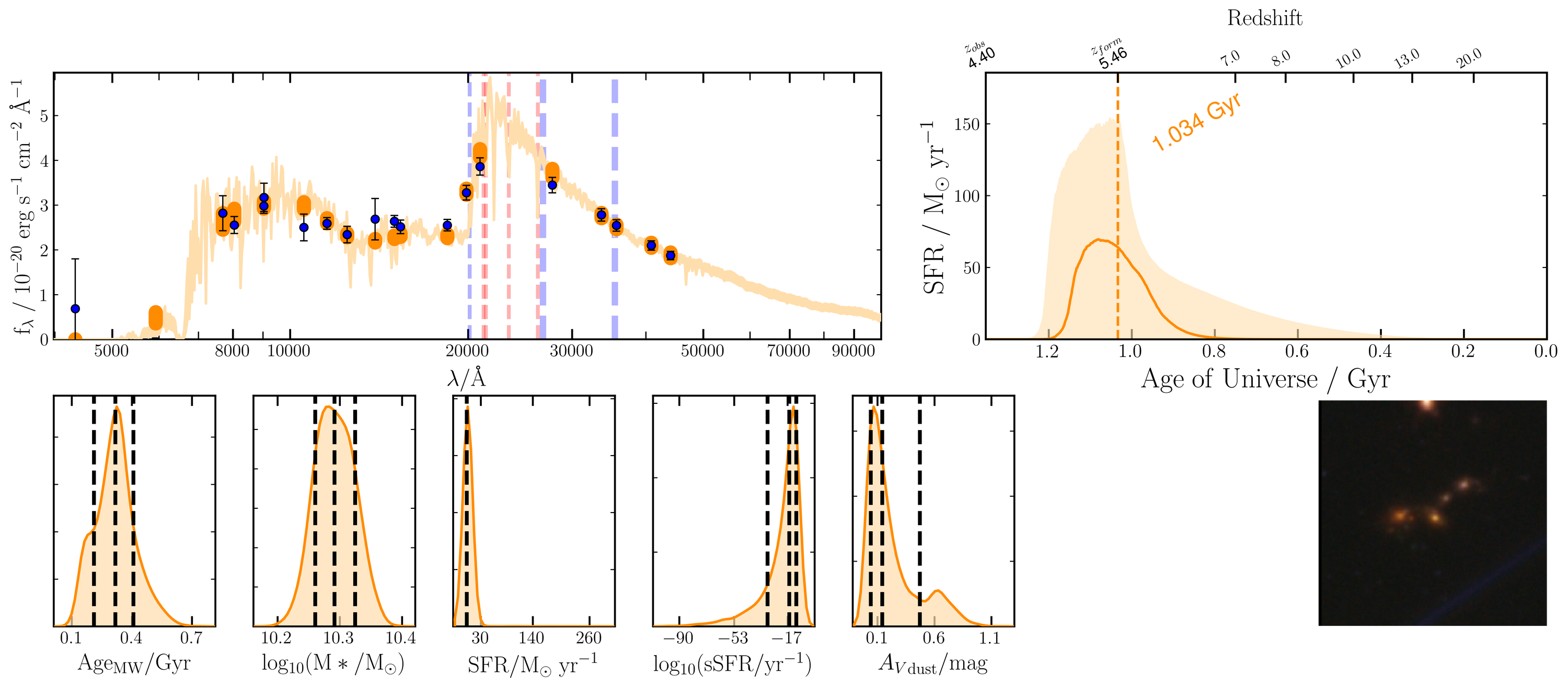}\\[1.5ex]

    \caption{SEDs of two spectroscopically confirmed quiescent galaxies (e.g., GS-21434 and GN-26273). The SED fitting is performed at fixed redshift $z_\mathrm{spec}$. Symbols and colors are as in Figure \ref{fig:combined-seds}.}
    \label{fig:spec-z-seds}
    
\end{figure*}

We cross-match our quiescent galaxy sample with the DAWN JWST Archive (DJA)\footnote{The data products presented herein were retrieved from the DAWN JWST Archive (DJA). DJA is an initiative of the Cosmic Dawn Center (DAWN), which is funded by the Danish National Research Foundation under grant DNRF140.} public NIRSpec datasets (v4.4) \citep{Brammer2022, Heintz2024, deGraaff2024}, and find that two of our photometric quiescent candidates have been spectroscopically confirmed. We perform SED fitting with the redshift fixed at their spectroscopic redshifts. The best-fit SEDs, SFHs, and cutouts are shown in Figure \ref{fig:spec-z-seds}.

The first galaxy, GS-21434, was initially identified as a quiescent candidate in a photometric study by \citet{Carnall2020}. It was later observed with JWST/NIRSpec PRISM in Cycle 1 (GO-2198; PIs: L. Barrufet $\&$ P. Oesch). The resulting spectrum confirmed a spectroscopic redshift of $z_\mathrm{spec} = 4.3559$ and showed no significant emission lines, confirming its quiescent nature \citep{Barrufet2025}. Its spectroscopic redshift places it at the edge of the z = 4.267 overdensity.

The second galaxy, GN-26273, was observed as part of the JADES spectroscopic follow-up program using NIRSpec with the G395H grating and F290LP filter. The spectrum confirmed a spectroscopic redshift of $z_\mathrm{spec} = 4.3997$ and 
revealed a faint H$\alpha$ emission line with a luminosity of $L_{\mathrm{H}\alpha} = 2.239 \times 10^{41}$ erg s$^{-1}$. We estimate the corresponding SFR using the calibration from \citet{Pflamm-Altenburg2007} for a \citet{Kroupa2001} IMF: 
\begin{equation}
\mathrm{SFR} = \frac{L_{\mathrm{H}\alpha}}{1.89 \times 10^{41} [\mathrm{erg} \ \mathrm{s}^{-1}]} [M_\odot \ \mathrm{yr}^{-1}]
\end{equation}

This yields an SFR of approximately $1.2 M_\odot\mathrm{yr}^{-1}$, which is consistent with a low level of star formation and with its quiescent classification.

This corresponds to $\log$(sSFR/yr$^{-1}$) $\sim -10.2$, still below our sSFR threshold of 0.2/t$_{obs}$ at $z = 4.40$, and higher than the SED-based value in Table \ref{table:quiescent-candidates}, which is averaged over 100 Myr and is poorly constrained at low SFR.

\section{Discussion} \label{sec:discussion}
Our study provides new insights into the properties of quiescent galaxies in proto-clusters at $z\sim4-6$. Although we do not observe a strong dependence of quiescent fraction on redshift, we find that the quiescent fraction generally increases with stellar mass (see Figure \ref{fig:quiescent-fraction}). In the highest mass bin ($\log$($M_*/M_{\odot}$)$>10.5$), this trend is particularly pronounced in the GOODS-N proto-clusters at $z=4.406$ and $z=5.194$, where the quiescent fraction reaches 100\% (2/2 and 1/1 respectively), corresponding to $\log_{10}$(f$_Q$ + 1)$\sim0.3$ in Figure \ref{fig:quiescent-fraction}. However, this mass dependence is not universal, and whether it holds generally in the early universe remains an open question. In the GOODS-S proto-clusters at $z\sim4.267$ and $z=5.384$, for example, no massive quiescent galaxy is observed in the same high-mass bin, highlighting significant proto-cluster to proto-cluster variation in the quiescent fraction at the high-mass end. 

This behavior is not unique to our study: substantial scatter in the quiescent fraction at the high mass end ($\log$($M_*/M_{\odot}$)$>10.5$) is also seen among studies of lower redshift ($z\sim2-4$) proto-clusters \citep{Edward2024, Forrest2024a, McConachie2022, Tanaka2024}, which report values of $\sim20\%$ at $z=2-2.5$, $\sim30\%$ at $z=3.3$, $\sim70\%$ at $z=3.37$, and $\sim40\%$ at $z=4$. In fact, \citet{McConachie2025} has shown that galaxies exhibit conformity in $z>3$ proto-clusters: elevated quiescent fractions are found when the central ultra-massive galaxies ($\log$($M_{*}/M_{\odot}$) $>$ 11 ; hereafter UMG) is itself quiescent, whereas, proto-clusters with star-forming UMGs show lower quiescent fractions, similar to the field. This evidence for galactic conformity at $z>3$ could help explain the large diversity of quiescent fractions seen among high redshift proto-clusters. This behavior can also be understood as a consequence of massive cluster galaxies undergoing more rapid, earlier formation and more efficient quenching via AGN feedback \citep{Casey2016}.

In contrast, in the intermediate mass bin ($9.75 < \log(M_*/M_{\odot})< 10.50$), our average quiescent fraction of $3.5\%$ across four proto-clusters is broadly consistent with these earlier works. 

To assess whether the proto-cluster environment enhances quiescence relative to the field at comparable epochs, we compare our results to field quiescent fraction measurements at z$\sim 4-6$ from \citet{Weaver2023}, \citet{Kakimoto2026}, \citet{Hegde2026}, \citet{Shuntov2026}, and \citet{Yang2026}. In the highest mass bin, the quiescent fractions of the two GOODS-N proto-clusters are higher than the field (Figure \ref{fig:quiescent-fraction}), whereas the GOODS-S proto-clusters show no such enhancement. This suggests that proto-cluster membership alone does not guarantee enhanced quenching of the most massive galaxies. Instead, this comparison is consistent with the galactic conformity scenario described above, in which the quenching state of the central UMG primarily governs the elevated quiescent fractions we observe. We note that these field measurements are based on different photometric catalogs, SED modeling, and selection criteria, so this comparison is qualitative.

We examine the location of quiescent galaxies relative to the peaks within the proto-clusters in Figures \ref{fig:gdn-overdensity-all} and \ref{fig:gds-overdensity-all} to further investigate the role of environment. First, the figures show that the proto-clusters are broken down into sub-components, suggesting dark matter halos at these high redshifts are not yet virialized. Furthermore, the quiescent galaxies tend to distribute themselves around the peaks of the proto-clusters, but do not usually coincide with the peaks. This observation agrees with recent results from Magneticum Pathfinder \citep{Kimmig2023}, which find quiescent galaxies at $z\sim3$ sit in the local underdensities rather than the densest nodes. They explain that these quiescent galaxies move towards under-dense regions during quenching, as structure growth creates new nodes around them \citep{Kimmig2023}. Alternatively, proto-clusters selected using line-emitting galaxies may not fully represent the overall galaxy population or underlying dark matter distribution, thus misidentifying the peak of the overdensity.

We also note that $75.0\%$ (6/8) of our robust quiescent galaxies -- and $66.7\%$ (8/12) of the total quiescent sample -- are associated with the $z\sim4.267$ and the $z=4.406$ proto-clusters, while we find four and no quiescent galaxies in the $z = 5.194$ and $z=5.384$ proto-cluster, respectively. Only one of our robust quiescent galaxies (GN-31416) lies at z $>$ 5. This scarcity of robust quiescent galaxies at $z \geq 5$ aligns with predictions from large-volume cosmological hydrodynamical simulations, including FLARES \citep{Lovell2023}, Magneticum Pathfinder \citep{Kimmig2023}, TNG300 \citep{Hartley2023}, as well as the semi-analytic model GAEA \citep{DeLucia2024}. These simulations predict a drastic decrease in the number density of quiescent galaxies with stellar masses $\log$(M$_*$/M$_{\odot}$) $> 10$ at $z\geq5$ \citep{Lovell2023}, with many simulations containing zero quiescent galaxies at $z>5$. The scarcity of massive quiescent galaxies beyond this redshift in both observations and simulations suggests that the mechanisms (e.g., AGN feedback) responsible for quenching star formation in massive galaxies are not yet prevalent in the early universe but emerge later in cosmic history. However, the presence of quiescent candidates in the $z = 5.194$ proto-cluster suggests that some galaxies in dense environments may have quenched by z $\sim$ 5, earlier than most simulations predict. Spectroscopic follow-up will be needed to confirm these candidates.

Our analysis has several limitations. The reliance on photometric redshifts and SED modeling introduces uncertainties, particularly in distinguishing high-$z$ dusty star-forming galaxies from quiescent galaxies \citep{Long2024, Antwi-Danso2023}. Moreover, the small area of the FRESCO survey (124 arcmin$^2$) limits our ability to generalize these findings across larger cosmic volumes. The selection of QGs at high redshift is not trivial. We implement multiple rest-frame color techniques and impose an sSFR cut from SED modeling to minimize the biases inherent in any single approach. We also allow for more generous criteria in rest-frame color selection to include younger quiescent galaxies. 

Nonetheless, our use of JWST/NIRCam data, coupled with robust selection techniques, provides a powerful tool for identifying and characterizing quiescent galaxies in the early universe. As shown earlier, our techniques can recover spectroscopically confirmed quiescent galaxies from \citet{Barrufet2025} and \citet{Baker2025}. Even the bluest, youngest quiescent galaxy in \citet{Baker2025}, which lies only $\sim$0.1 mag below our UVJ boundary, is consistent with our selection within its 1$\sigma$ color uncertainties. The inclusion of grism spectroscopy to confirm the absence of emission lines, except for the z $\sim$ 4.267 overdensity, strengthens our confidence in our quiescent classifications.

Future JWST/NIRSpec spectroscopic follow-up will be essential to validate and expand upon these findings. Such efforts will confirm redshifts and the quiescent nature of galaxies, refine stellar mass and SFR estimates, and verify the existence of structures such as the $z\sim 4.267$ proto-cluster in GOODS-S. Additionally, a more robust comparison between the quiescent fractions in proto-clusters and in the field (non-overdense regions) would require selecting field quiescent galaxies with the same photometric catalogs, SED modeling, and selection criteria used here over a larger area, which is beyond the scope of this paper. Large medium-band JWST programs, such as MINERVA \citep{Muzzin2025} and SPAM (R. Larson, K. Davis et al. 2026, in preparation), will extend studies of high-redshift proto-clusters and their quiescent members, providing the spectral resolution and completeness over a much wider surveyed area necessary to enable more robust measurements of quiescent fractions both within proto-clusters and in the field, disentangling the effect of environment on quiescent galaxies at high redshift.

\renewcommand{\arraystretch}{1.1}
\begin{deluxetable*}{ccccccccccc}
\tablecaption{The list of our final quiescent candidates with the selection criteria they satisfied and their important physical parameters including stellar masses $\log$(M$_*$/M$_{\odot}$), redshift $z$, star-formation-rate log(SFR), specific star-formation rate $\log$(sSFR), and mass-weighted age Age$_{MW}$. The IDs of robust quiescent candidates are shown in bold. \label{table:quiescent-candidates}}
\tablehead{
\colhead{ID} & \colhead{M1} & \colhead{M2} & \colhead{M3} & \colhead{M4} & \colhead{M5} & \colhead{$\log$(M$_*$/M$_{\odot}$)} & \colhead{$z$} & \colhead{$\log$(SFR/M$_{\odot}$ yr$^{-1}$)} & \colhead{$\log$(sSFR/yr)} & \colhead{Age$_{MW}$/Gyr}
}
\startdata
\textbf{GN-31537} & v & v & v & v & v & 11.13$_{-0.02}^{+0.02}$ & 4.41$_{-0.03}^{+0.02}$ & 0.66$_{-0.13}^{+0.16}$ & -10.46$_{-0.13}^{+0.15}$ & 0.53$_{-0.05}^{+0.10}$ \\
\textbf{GN-15147} & v & v & v & v & v & 10.71$_{-0.06}^{+0.05}$ & 4.28$_{-0.14}^{+0.10}$ & 0.05$_{-0.00}^{+1.74}$ & -10.68$_{-32.44}^{+1.73}$ & 0.69$_{-0.26}^{+0.25}$\\ 
GN-5896 & v &  &  &  & v & 10.71$_{-0.07}^{+0.07}$ & 5.32$_{-0.12}^{+0.13}$ & 0.41$_{-0.21}^{+0.22}$ & -10.31$_{-0.16}^{+0.17}$ & 0.90$_{-0.07}^{+0.05}$ \\ 
\textbf{GN-31422} & v & v & v & v & v & 10.38$_{-0.03}^{+0.02}$ & 4.39$_{-0.03}^{+0.02}$ & -10.20$_{-0.00}^{+8.92}$ & -20.57$_{-36.93}^{+8.91}$ & 0.81$_{-0.15}^{+0.05}$ \\
\textbf{GN-26273} & v & v & v & v & v & 10.29$_{-0.03}^{+0.03}$ & 4.3997 & -5.63$_{-14.61}^{+4.71}$ & -15.93$_{-14.63}^{+4.72}$ & 0.32$_{-0.11}^{+0.09}$ \\ 
\textbf{GN-31416} & v & v & v & v & v & 10.2$_{-0.06}^{+0.06}$ & 5.57$_{-0.23}^{+0.15}$ & -1.91$_{-15.85}^{+3.55}$ & -12.11$_{-15.74}^{+3.47}$ & 0.23$_{-0.09}^{+0.11}$ \\ 
\textbf{GN-20113} & v & v & v & v & v & 9.22$_{-0.04}^{+0.05}$ & 4.89$_{-0.10}^{+0.15}$ & -7.06$_{-0.00}^{+5.76}$ & -16.27$_{-28.33}^{+5.73}$ & 0.35$_{-0.09}^{+0.14}$ \\ 
GN-28956 & v & v &  & v & v & 9.07$_{-0.04}^{+0.04}$ & 5.10$_{-0.14}^{+0.10}$ & -0.56$_{-5.21}^{+0.86}$ & -9.63$_{-5.19}^{+0.83}$ & 0.17$_{-0.03}^{+0.06}$ \\  
GN-31274 & v & v &  & v & v & 9.04$_{-0.04}^{+0.03}$ & 4.52$_{-0.05}^{+0.07}$ & -3.46$_{-10.5}^{+2.85}$ & -12.49$_{-10.53}^{+2.84}$ & 0.20$_{-0.04}^{+0.05}$ \\
\textbf{GS-21434} & v & v & v & v & v & 10.37$_{-0.02}^{+0.03}$ & 4.3559 & -19.33$_{-53.38}^{+16.02}$ & -29.67$_{-53.42}^{+15.97}$ & 0.56$_{-0.08}^{+0.13}$ \\
\textbf{GS-10005} & v & v & v & v & v & 8.98$_{-0.04}^{+0.04}$ & 4.30$_{-0.05}^{+0.03}$ & -1.13$_{-0.70}^{+0.31}$ & -10.11$_{-0.66}^{+0.30}$ & 0.71$_{-0.21}^{+0.19}$ \\ 
GS-29940 & v & v &  &  & v & 8.95$_{-0.05}^{+0.04}$ & 4.07$_{-0.09}^{+0.06}$ & -3.05$_{-8.82}^{+2.53}$ & -12.02$_{-8.75}^{+2.54}$ & 0.20$_{-0.03}^{+0.04}$ \\ 
\enddata

\tablecomments{
M1: selection by rest-frame colors UVJ described in section \ref{sec:uvj}. \\
M2: selection by synthetic rest-frame colors (ugi)$_s$ described in section \ref{sec:ugi}. \\
M3: selection by Gaussian Mixture Model (GMM) trained on rest-frame NUVU-VJ colors with quiescent probability bigger than 10\%, or p(q)$>$ 0.1 described in section \ref{sec:nuvu-vj}. \\
M4: selection by Gaussian Mixture Model (GMM) trained on rest-frame NUVU-VJ colors with quiescent probability bigger than 1\%, or p(q)$>$ 0.01 described in section \ref{sec:nuvu-vj}. \\
M5: selection by low specific star-formation rate i.e. sSFR $\leq$ 0.2/t$_{obs}$ at the time of observation given BAGPIPES SED modeling described in section \ref{sec:low-ssfr}. \\
For galaxies whose SFH posteriors extend to SFR $\rightarrow$ 0, $\log$(SFR) and $\log$(sSFR) are unbounded below; values below $\log$(sSFR/$yr^{-1}$) = --11 should be interpreted as consistent with no ongoing star formation (as in Figure \ref{fig:ssfr-mass}).
}
\end{deluxetable*}

\begin{acknowledgments} \label{sec:acknowledgments}
Support for program JWST-GO-01895 was provided by NASA through a grant from the Space Telescope Science Institute, which is operated by the Association of Universities for Research in Astronomy, Incorporated, under NASA contract NAS5-26555. We acknowledge support from the FRESCO collaboration, which made this work possible, as well as the immense help from colleagues in the Physics and Astronomy Department at Tufts University.

AI tools were used to suggest language improvements and to check LaTeX formatting errors. The GAI tool used was Claude Opus 5.5 (Anthropic).

\end{acknowledgments}

\facilities{JWST(NIRCam, MIRI), HST(ACS, WFC3)}

\software{BAGPIPES \citep{Carnall2018}, 
          EAZY \citep{Brammer2008}, 
          CLOUDY \citep{ferland2013}, 
          Astropy \citep{astropy:2013, astropy:2018, astropy:2022}}

\clearpage 

\appendix
\section{Additional figures and table}\label{app:sed}

\figsetstart
\figsetnum{9}
\figsettitle{SED, SFH, and cutout of all quiescent galaxies} 

\figsetgrpstart
\figsetgrpnum{9.1}
\figsetgrptitle{GN-31537}
\figsetplot{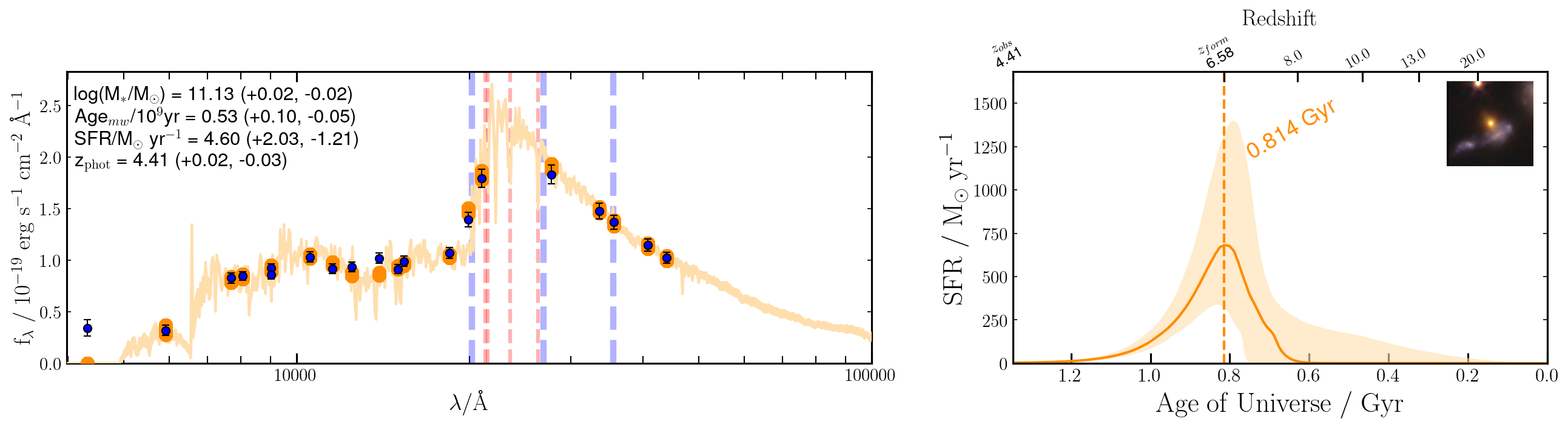}
\figsetgrpend

\figsetgrpstart
\figsetgrpnum{9.2}
\figsetgrptitle{GN-15147}
\figsetplot{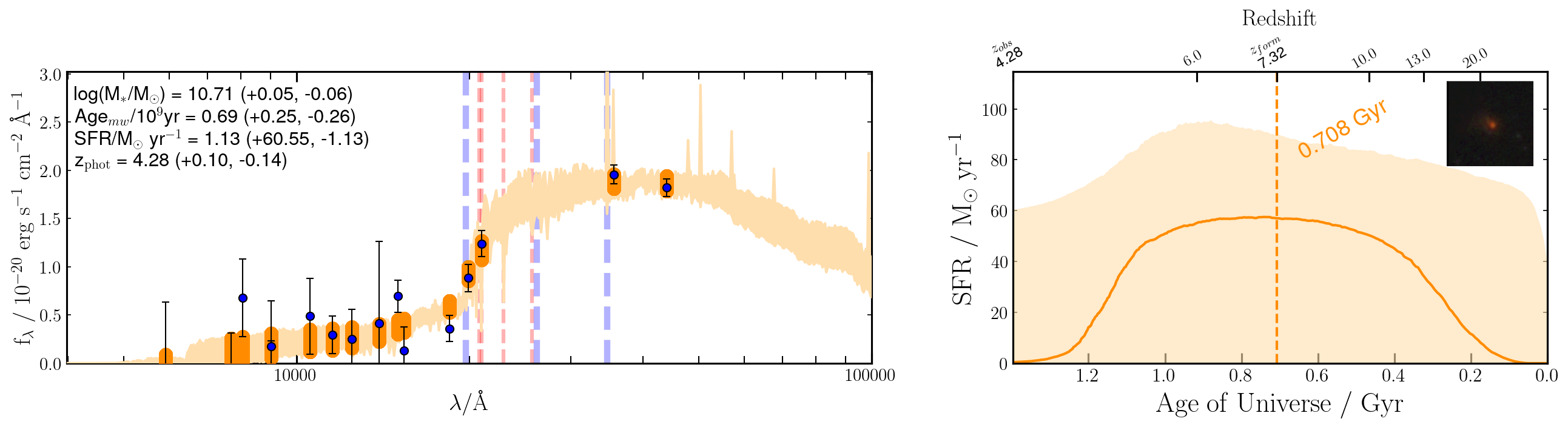}
\figsetgrpend

\figsetgrpstart
\figsetgrpnum{9.3}
\figsetgrptitle{GN-5896}
\figsetplot{GN-5896_appendix.png}
\figsetgrpend

\figsetgrpstart
\figsetgrpnum{9.4}
\figsetgrptitle{GN-31422}
\figsetplot{GN-31422_appendix.png}
\figsetgrpend

\figsetgrpstart
\figsetgrpnum{9.5}
\figsetgrptitle{GN-26273}
\figsetplot{GN-26273_appendix.png}
\figsetgrpend

\figsetgrpstart
\figsetgrpnum{9.6}
\figsetgrptitle{GN-31416}
\figsetplot{GN-31416_appendix.png}
\figsetgrpend

\figsetgrpstart
\figsetgrpnum{9.7}
\figsetgrptitle{GN-20113}
\figsetplot{GN-20113_appendix.png}
\figsetgrpend

\figsetgrpstart
\figsetgrpnum{9.8}
\figsetgrptitle{GN-28956}
\figsetplot{GN-28956_appendix.png}
\figsetgrpend

\figsetgrpstart
\figsetgrpnum{9.9}
\figsetgrptitle{GN-31274}
\figsetplot{GN-31274_appendix.png}
\figsetgrpend

\figsetgrpstart
\figsetgrpnum{9.10}
\figsetgrptitle{GS-21434}
\figsetplot{GS-21434_appendix.png}
\figsetgrpend

\figsetgrpstart
\figsetgrpnum{9.11}
\figsetgrptitle{GS-10005}
\figsetplot{GS-10005_appendix.png}
\figsetgrpend

\figsetgrpstart
\figsetgrpnum{9.12}
\figsetgrptitle{GS-29940}
\figsetplot{GS-29940_appendix.png}
\figsetgrpend

\figsetend

\begin{figure*}[ht!]
    \centering
    \vspace{1em}
    \textbf{GN-31537}\\
    \vspace{-1em}
    \includegraphics[width=\textwidth]{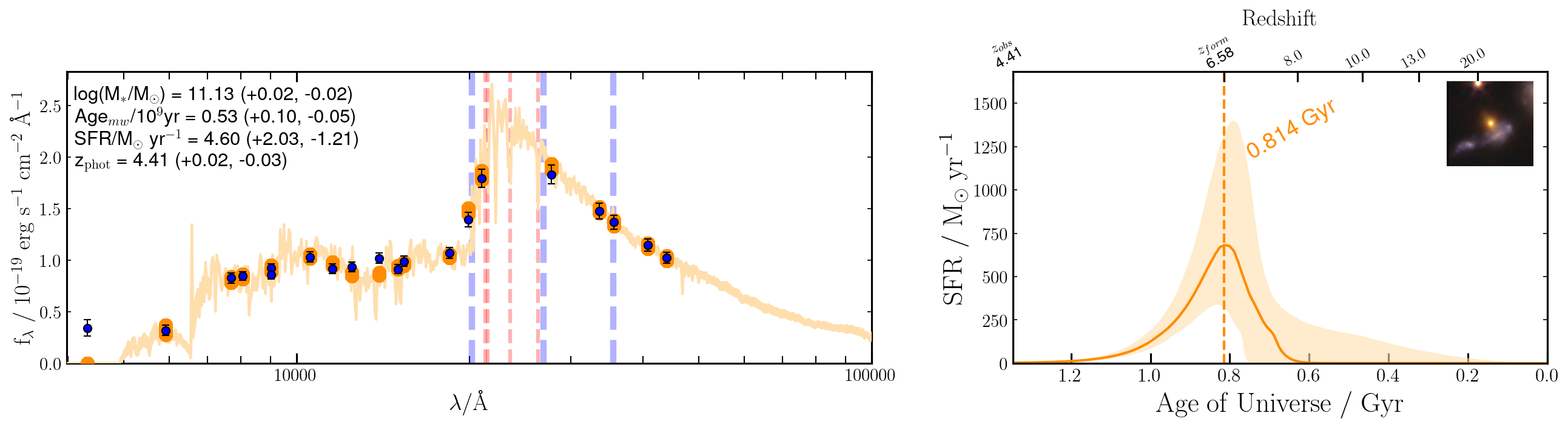}\\[2ex]

    \textbf{GN-15147}\\
    \vspace{-1em}
    \includegraphics[width=\textwidth]{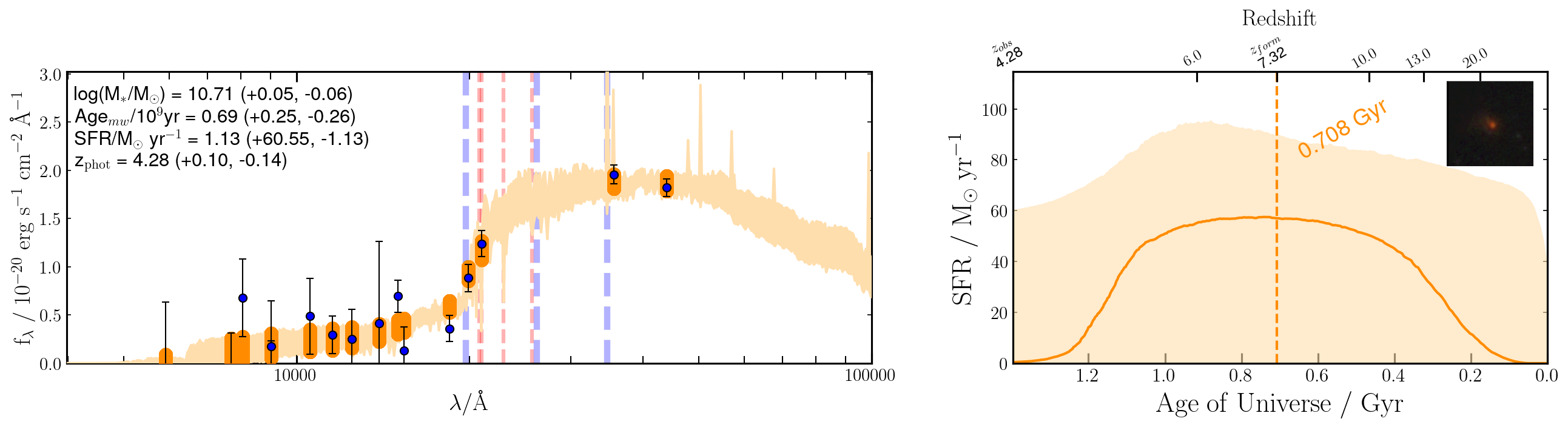}
    
    \caption{Supplemental SED modeling diagnostics for all candidate quiescent galaxies in our sample. The complete figure set (12 images) is available in the online journal. Observed data are shown in blue, and the posterior SEDs from BAGPIPES are shown in orange. For galaxies with photometric redshifts, the SED fitting is performed with the redshift prior using the 16th and 84th percentile of EAZY’s p(z). For galaxies with spectroscopic redshifts, the redshift is fixed at the spectroscopic redshift. The vertical blue dashed lines represent emission features, including [OII], [OIII], and H$\alpha$ + [NII]. The red dashed lines represent absorption features, including CaH+K, H$\beta$, and H$\gamma$. Star-formation history shows the star-formation activity from the Big Bang (right) to the time when the galaxy is observed (left). The vertical orange dashed line is $t_{form}$, when the galaxy formed $50\%$ of its mass. The RGB cutouts are composed by F182M, F210M, and F444W bands with a size of $2'' \times 2''$. }
    \label{fig:appendix-seds}
\end{figure*}
\clearpage

\begin{figure*}[h]
    \centering
    \includegraphics[width=.9\textwidth]{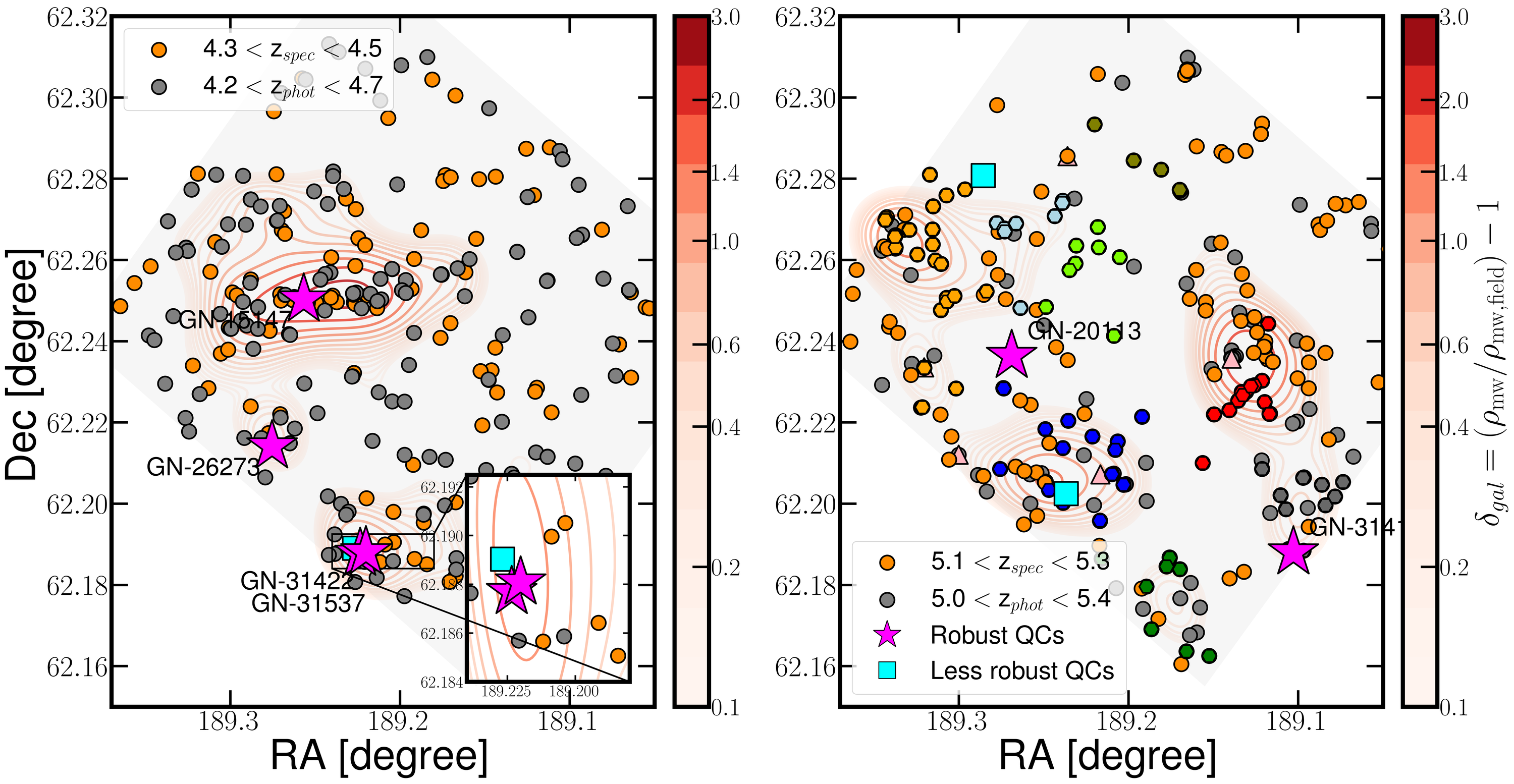}
    \includegraphics[width=.9\textwidth]{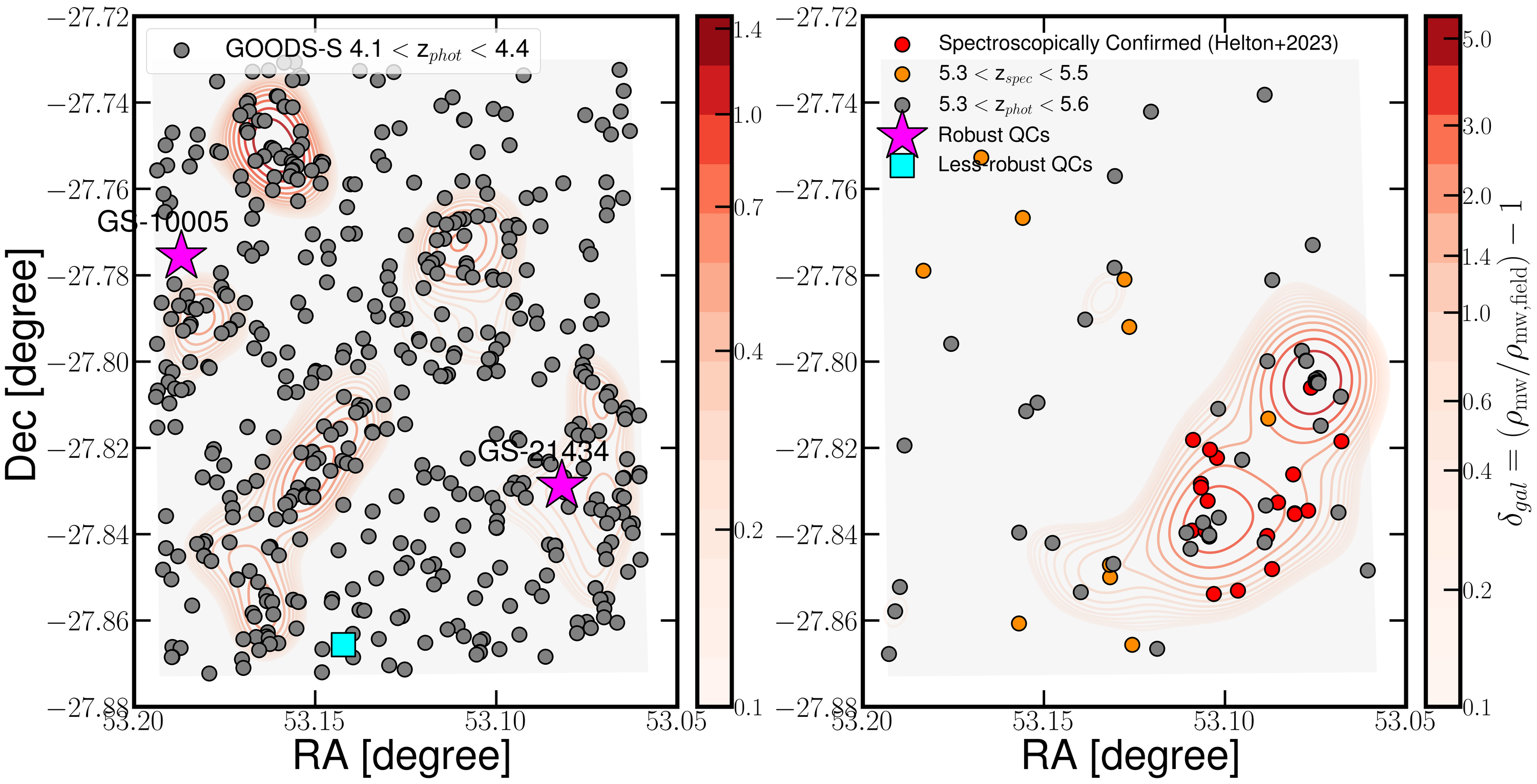}
    \caption{The contour line represents the galaxy overdensities in the proto-clusters weighted by stellar masses. The mass-weighted overdensity contrast is defined as:  $\delta_{\rm gal, mw} = (\rho_{mw} /\rho_{\rm mw, field}) - 1$, where the mass-weighted number density $\rho_{mw}$  is defined as $\rho_{mw} = \Sigma \log (M_*/10^6 M_{\odot})/Area$.  The top panels show the proto-clusters in GOODS-N at $z=4.406$ (left) and $z=5.194$ (right). The bottom panels show the proto-clusters in GOODS-S at $z=4.267$ (left) and $z=5.384$ (right). Symbols and colors are the same as in Figure \ref{fig:gdn-overdensity-all}}
    \label{fig:overdensity-all-mweighted}
\end{figure*}

\begin{figure*}[h]
    \centering
    \includegraphics[width=.48\textwidth]{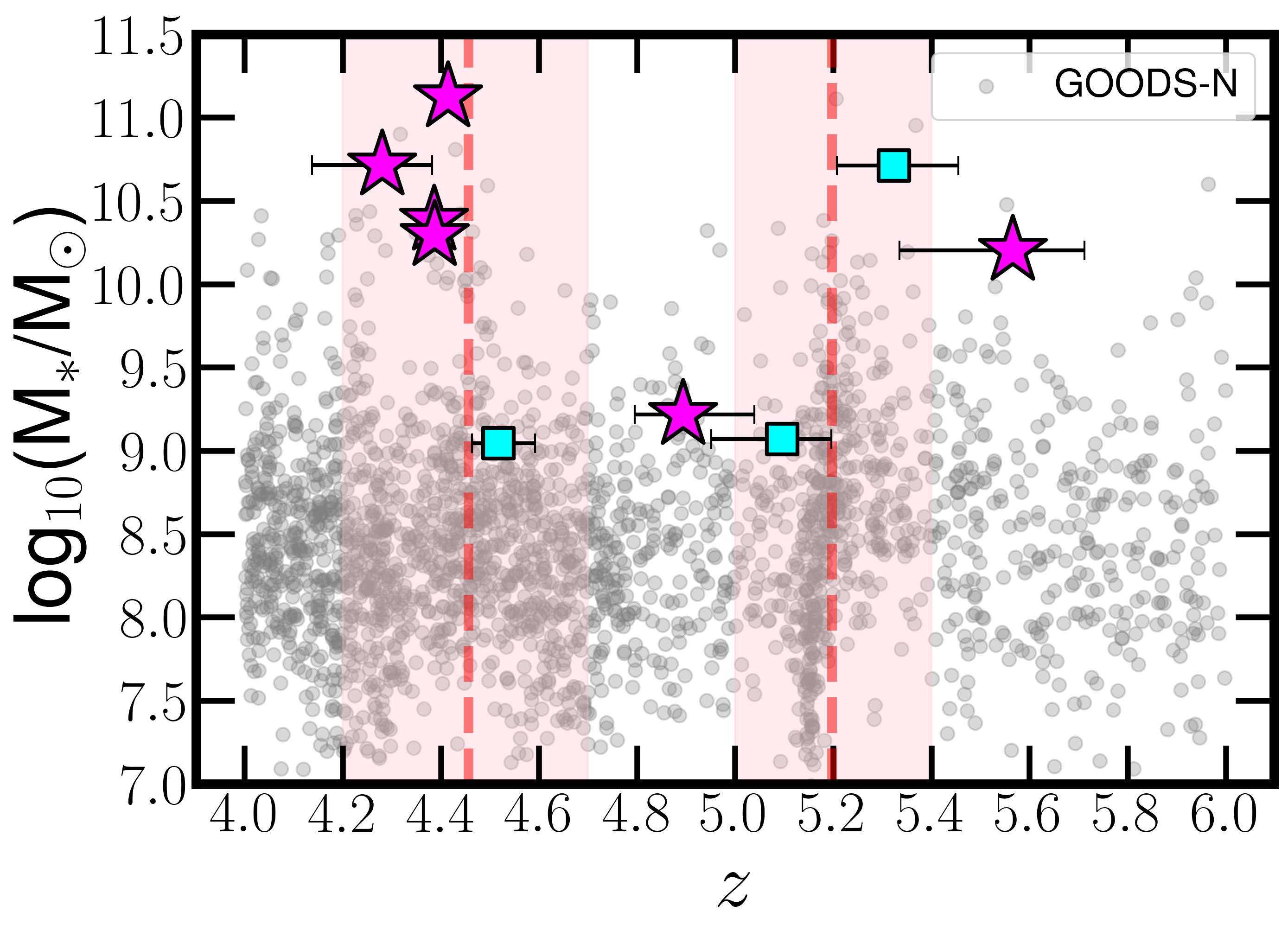}%
    \hfill
    \includegraphics[width=.48\textwidth]{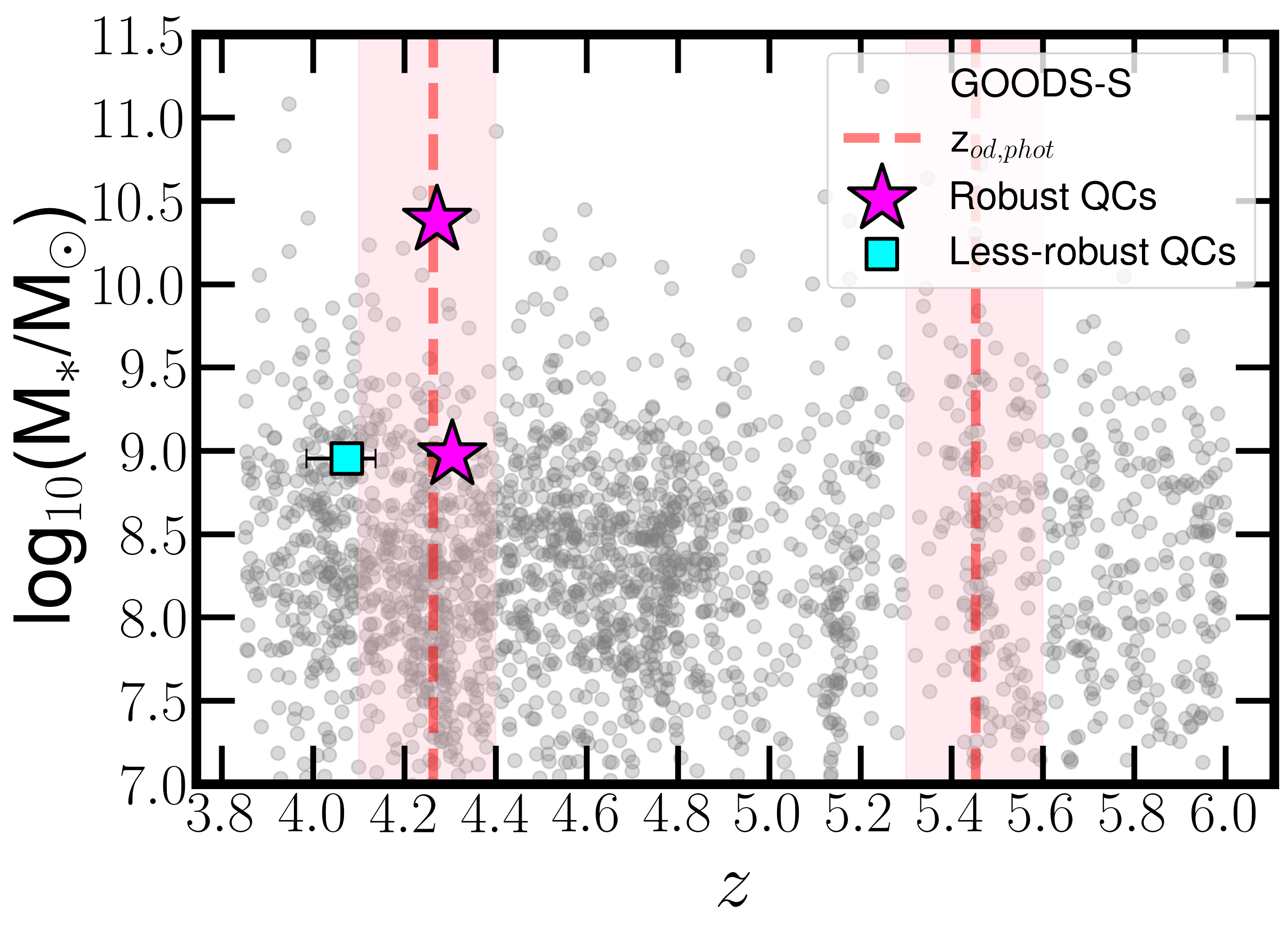}
    \caption{Stellar mass vs redshift for galaxies at $z = 4 - 6$ in GOODS-N (left) and GOODS-S (right). Robust quiescent galaxies are shown in magenta stars. Other quiescent candidates are shown in cyan squares. The vertical red lines represent the photometric redshifts of the four proto-clusters: $z_{\rm od, phot} = 4.456$, $5.197$, $4.267$, and $5.452$ from left to right. The pink shaded area represents the range of photometric redshift that used to select quiescent galaxies corresponding to each proto-cluster.}
    \label{fig:mass-z}
\end{figure*}

\renewcommand{\arraystretch}{1.1}
\begin{deluxetable*}{lcc}
    \tabletypesize{\normalsize}
    \tablewidth{\textwidth}
    \tablecaption{Coordinates of Quiescent Galaxy Candidates\label{table:coordinates}}
    \tablehead{
        \colhead{ID} &
        \colhead{R.A.} &
        \colhead{Decl.} \\
        \colhead{} &
        \colhead{(deg)} &
        \colhead{(deg)}
    }
    \startdata
    GN-31537  & 189.22349 &  62.18769 \\
    GN-15147  & 189.25677 &  62.25030 \\
    GN-5896  & 189.28554 &  62.28078 \\
    GN-31422  & 189.22014 &  62.18808 \\
    GN-26273  & 189.27544 &  62.21414 \\
    GN-31416  & 189.10285 &  62.18805 \\
    GN-20113  & 189.26881 &  62.23640 \\
    GN-28956  & 189.23670 &  62.20247 \\
    GN-31274  & 189.22677 &  62.18903\\
    GS-21434  &  53.08188 & -27.82879 \\
    GS-10005  &  53.18697 & -27.77562 \\
    GS-29940  &  53.14217 & -27.86564 \\
    \enddata
    \tablecomments{Coordinates are given in J2000 equatorial coordinates in decimal degrees.}
\end{deluxetable*}

\clearpage
\bibliography{main}{}
\bibliographystyle{aasjournal}

\end{document}